\documentclass[prd,aps,
superscriptaddress,preprintnumbers,nofootinbib,showpacs]{revtex4}

\usepackage{graphicx}
\usepackage{xcolor}
\usepackage{exscale}
\usepackage[intlimits]{amsmath}
\usepackage{amsfonts}
\usepackage{amssymb,amscd}
\usepackage{epsfig}
\usepackage[english]{babel}
\usepackage{amsfonts,amsmath,amssymb}
\usepackage{booktabs}
\usepackage{float,braket}
\usepackage{pslatex}
\usepackage{graphicx,color}
\usepackage{subfigure}
\usepackage{graphicx,wrapfig}
\usepackage[utf8]{inputenc}

\newcommand{\beq}{\begin{equation}}
\newcommand{\eeq}{\end{equation}}

\newcommand{\vk}{\vec{k}}

\newcommand{\mG}{\mathcal{G}}
\newcommand{\mM}{\mathcal{M}}

\numberwithin{equation}{section}

\newcommand{\half}{{\textstyle{\frac 1 2}}}

\newcommand{\drop}[1]{}

\newcommand{\p}{\partial}

\newcommand{\beqa}{\begin{eqnarray}}
\newcommand{\eeqa}{\end{eqnarray}}

\newcommand{\be}{\begin{equation}}
\newcommand{\ee}{\end{equation}}
\renewcommand{\eqref}[1]{Eq.~(\ref{#1})}

\renewcommand{\L}{\ensuremath{\mathcal L}}

\begin{document}

\title{Schwinger-Dyson Equations in Coulomb Gauge Consistent with Numerical Simulation}

\author{Patrick~Cooper}
\email[]{cooperp@duq.edu}
\affiliation{Duquesne University, Pittsburgh, PA 15282}

\author{Daniel~Zwanziger}
\email[]{dz2@nyu.edu}
\affiliation{New York University, New York, NY 10003}

\date{October 13, 2018}

\begin{abstract} 
 
In the present work we undertake a study of the Schwinger-Dyson equation (SDE)
in the Euclidean formulation of local quantum gauge field theory, with Coulomb
gauge condition $\p_i A_i = 0$. We continue a previous study  which kept only
instantaneous terms in the SDE that are proportional to $\delta(t)$ in order to
calculate the instantaneous part of the time component of the gluon propagator
$D_{A_0 A_0}(t, R)$.  We compare the results of that study with a numerical
simulation of lattice gauge theory and find that the infrared critical
exponents and related quantities agree to within 1\% to 3\%.  This raises the
question, ``Why is the agreement so good, despite the systematic neglect of
non-instantaneous terms?" We discovered the happy circumstance that all the
non-instantaneous terms are in fact zero.  They are forbidden by the symmetry
of the local action in Coulomb gauge under time-dependent gauge transformations
$g(t)$. This remnant gauge symmetry is not fixed by the Coulomb gauge
condition. The numerical result of the present calculation is the same as in
the previous study; the novelty is that we now demonstrate that all the
non-instantaneous terms in the SDE vanish.  We derive some elementary
properties of propagators which are a consequence of the remnant gauge
symmetry.  Our results support the simple physical scenario
in which confinement is the result of a linearly rising color-Coulomb
potential, $V(R) \sim \sigma R$ at large $R$.  We also show that the horizon condition $\langle H(gA) \rangle = (N^2 -1) d V$, and the divergence of the ghost dressing function at ${\bf k} = 0$, $\lim_{|{\bf k} \to 0|}{\bf k}^2 D_{c \bar c}({\bf k}) = \infty$, are identical gauge conditions. \\
\end{abstract}

\maketitle

\section{Introduction}








While the quest for exotic quantum theories of gravity captivates many
physicists, a much more mundane question remains unanswered: what is the
qualitative mechanism for the mismatch between the UV degrees of freedom of the
standard model (quarks and gluons) and the IR states we observe in the lab
(baryons and mesons).  In other words, an intuitive physical picture of
confinement still eludes us, despite the empirical successes of the standard
model in the UV. Genuinely new physics is unlikely needed; from lattice
simulations, we know that non-Abelian gauge theory by itself is capable of creating
gluonic flux tubes which confine quark-anti-quark pairs into mesons at low
energy \cite{Teper:2009uf}. Yet despite our best efforts, the mathematics behind this
phenomenon is unknown. The ultimate goal of science is not just to
reproduce nature, but rather to understand it, and this goal is what drives the
field of non-perturbative QCD.  

The breakdown of perturbation theory at low energies forces one to face
the non-Abelian character of Yang-Mills theory head on. Various approaches have been
made over the years to use functional methods to extract information about the
fully non-perturbative, dressed propagators and vertices of QCD. These
quantities are crucial to understanding confinement. For example, an infrared
vanishing gluon propagator violates reflection positivity and thus implies that
the gluon is not an asymptotic field of the theory. Also, in Landau gauge, the
divergence of the ghost dressing function at $k = 0$ leads to a well defined global color
charge which is an important part of the Kugo-Ojima confinement scenario
\cite{Fischer:2006vf, Kugo01021979}. Two techniques that have greatly increased
our understanding of the non-perturbative sector of QCD, constituting an
infinite hierarchy of coupled equations that can be derived rigorously from the
full quantum effective action, are the functional renormalization group
equations (FRG) \cite{Weber:2011,Reinhardt:2010xm, Fischer:2006vf,
Fischer:2009tn} and the Schwinger-Dyson equations (SDE) \cite{Blum:2015lsa,
Vujinovic:2014fza, Watson:2011nn, Watson:2010cn, Huber:2010cq, Huber:2009tx,
Lichtenegger:2009dw, Alkofer:2009dm}. A third technique, exploits a formal
similarity between vacuum expectation values in the Hamiltonian formalism and
correlation functions in Euclidean quantum field theory. In this approach an
ansatz is made for the vacuum wave functional which confirms results found by other
techniques \cite{Campagnari:2015zsa, Huber:2014isa, Campagnari:2013hn,
Reinhardt:2011hq, Weber:2011, Epple:2006hv, Schleifenbaum:2006bq}. The
advantage of the canonical approach is that with Lagrangian methods, an
uncontrolled truncation must be made to complete the equations. At first
glance, it seems that in the Hamiltonian approach, a truncation is still made
even with a non-Gaussian ansatz: a finite order polynomial is still used for
the vacuum wave functional.  However, due to the gap equation found by varying
the energy density, the best possible coefficients of that finite order
polynomial will be found which minimizes the effect of the truncation
\cite{Campagnari:2015zsa}.  Nonetheless, we will proceed with the approach
povided by the Schwinger-Dyson equations.  In contradistinction to the
Hamiltonian operator method, we use a local Euclidean quantum field theory.

Coulomb gauge is a natural choice for attempting a qualitative understanding of
confinement for two reasons. First, it is a unitary gauge, where Gauss's law
can be resolved explicitly by the longitudinal component of the color electric
field, thus only propagating physical degrees of freedom (analogous to the two
polarizations of the physical photon of QED). To interpolate between UV QCD and
phenomenological theories of IR QCD, tracking the physical degrees of freedom
is essential. Second, the long-range nature of the color-Coulomb potential,
$\delta(x_0 - y_0) V_{Coulomb}(\vec{x} - \vec{y}) = \langle A_0(x) A_0(y) \rangle$,
gives a physical picture of what does the confining. Despite
being a gauge-dependent quantity, the color-Coulomb potential also gives us
insight into the IR asymptotics of the gauge-invariant Wilson potential by the
following argument, found in detail in \cite{Greensite:2004ke}. Consider a
quark-anti-quark pair at separated points, $\vec{x}$ and $\vec{y}$ with $R
\equiv |\vec{x}-\vec{y}|$.  The correlator of two Wilson lines, $G(R,T)$,
extending an amount $T$ in the time direction is related to the Hamiltonian and
the state $|\psi_{\bar{q} q}\rangle$ by
\begin{eqnarray}
        G(R,T) &= \langle \frac{1}{2} Tr[L^{\dagger}(\vec{x},0,T) L(\vec{y},0,T)]\rangle \\
               &= \langle \psi_{\bar{q} q} | e^{-(H-E_0)T} | \psi_{\bar{q} q} \rangle
\end{eqnarray}
where $L(\vec{x},0,T)$ is a Wilson line extending from $0$ to $T$ at point
$\vec{x}$. Defining the logarithmic derivative, 
\begin{equation}
        V(R,T) = - \frac{d}{dT} \mathrm{log}[G(R,T)]
\end{equation}
one can show that the Coulomb energy is obtained in the limit $T \rightarrow
0$, and the energy of the flux tube ground state is obtained in the opposite
limit, $T \rightarrow \infty$. Since the latter is the ground state, at large
R (so one can neglect the self-energy contribution), $V_{Coulomb} = V(R,0) >
V(R,\infty) = V_{Wilson}$.  Thus the Coulomb potential must be at least linear
(possibly super-linear) in order to reproduce a linearly rising Wilson
potential like the one seen on the lattice. While a long-range Coulomb
potential is a necessary condition for confinement, it isn't a sufficient one.
Similarly to how charges screen each other to make neutral molecules despite
the presence of the long-range Coulomb potential, the QCD vacuum 
creates quark-anti-quark pairs, confining color charge despite the presence of
a long range color-Coulomb potential. Thus, even at high temperature, above the
deconfinement phase transition, the long range Coulomb force is present as seen
in \cite{Lichtenegger:2009dw}. 

The instantaneous character of the dynamics is of particular importance to
those interested in studying the so-called quark-gluon plasma at high
temperature. The common wisdom is that at high temperatures, typical momentum
transfer is large, and thus, due to asymptotic freedom, quarks and gluons will
behave like a weakly interacting plasma.  The presence of a long range
color-Coloumb potential at high temperature challenges this view, and suggests
that one might expect a strongly interacting fluid, despite the approximate
Stefan-Boltzmann like behavior witnessed by Karsch et al on the lattice
\cite{Karsch:2003jg}. This isn't contridictory with the renormalization group;
recall that in Coulomb gauge, the physical quantity $g^2 D_{A_0 A_0}$ is a
renormalization-group invariant \cite{Zwanziger:1998ez}. This phenomenology
would be similar to $\mathcal{N}=4$ super Yang-Mills (SYM) in the planar limit
as pointed out in \cite{Arnold:2007pg}.  The comparison of high temperature QCD
to $\mathcal{N}=4$ SYM, a strongly-coupled integrable theory, is particularly
intriguing in light of an article by Dubovsky and Gorbenko
\cite{Dubovsky:2015zey} which suggests that at large $N$, the theory of QCD
flux tubes may also be integrable, evading the no-go theorem in
\cite{Cooper:2014noa} by possessing a massless pseudoscalar mode in addition to
the usual goldstone modes of a string-like flux tube embedded in spacetime. If
the color-Coulomb potential is indeed stronger at high temperature than at zero
temperature, as the lattice calculation suggests \cite{Greensite:2004ke}, this
would imply that gluons are \emph{more} likely to form color singlets (ie.
glueballs), rather than less, since gluon configurations not bound into their
flux-tube ground state would be Boltzmann suppressed, making the flux tube
description more relevant. The instantaneous character of the dynamics is
crucial to accessing the physics at high temperature because it only keeps
terms in correlation functions that dominate at vanishingly small temporal
separation.  At high temperatures, the partition function becomes vanishingly
small in the Euclidean-time direction, thus yielding a dimensionally reduced
theory, in addition to any instantaneous physics inherited from the higher
dimensional theory. This heuristic picture is illustrated in
\cite{Arnold:2007pg} and a rigorous treatment of Gribov-Zwanger theory in
Coulomb gauge at finite temperature can be found in \cite{Cooper:2015bia}.

One objective of this article is to gain a quantitative handle on the
asymptotic behavior of the color-Coulomb potential. We do this by finding a
self-consistent set of vertices of the full quantum effective action that
satisfy the Schwinger-Dyson equations, continuing the work of
\cite{Alkofer:2009dm}. More specifically, in that work, only terms in the SDEs
were kept that are proportional to $\delta(t)$ in order to calculate the
instantaneous part of $D_{A_0 A_0}$, a.k.a., the color-Coulomb potential.  In
sect.\ \ref{conclusion} of the present article, we compare the infrared
critical exponents found \cite{Alkofer:2009dm} with numerical simulation in
lattice gauge theory of $SU(2)$ by Langfeld and Moyaerts
\cite{Langfeld:2004qs}.  The agreement is striking.  There is also reasonably
good agreement with Burgio, Quandt and Reinhardt \cite{Burgio:2012bk} for
$SU(2)$, and with Nakagawa et al \cite{Nakagawa:2009zf} for $SU(3)$. This led
us to question why the agreement was so good, in view of the neglect of the
non-instantaneous terms. We have discovered that the non-instantaneous terms
vanish because of the invariance under time-dependent gauge transformations
$g(t)$. These form the remnant gauge symmetry group of gauge transformations
that are not fixed by the Coulomb gauge condition $\p_i A_i = 0$.

\section{Local On-Shell Faddeev-Popov action in Coulomb gauge}

The Faddeev-Popov quantization of Yang-Mills theory in Coulomb gauge is defined
in phase-space formalism by the Lagrangian density,
\begin{align}
 \L^{\rm FP} &=  i \pi_i ( D_0 A_i - \p_i A_0) + \half \pi_i^2 + (1/4) F_{i j}^2  - \partial_i\bar c\cdot D_i c + i \p_i b \cdot A_i  \,,
\label{ko1_lag}
	\end{align}
where $F_{\mu \nu} = \p_\mu A_\nu - \p_\nu A_\mu + A_\mu\times A_\nu$ is the
Yang-Mills field strength \cite{Zwanziger:1998ez}. 
\normalsize
The connection $A^a_\mu$ as
well as the Nakanishi-Lautrup and Faddev-Popov ghost fields $b^a, c^a$ and
$\bar c^a$ are all fields in  the adjoint representation of the global  $SU(N)$
color group. Color components are represented by Latin superscripts. To
streamline notation we adopt the convention that $X\cdot Y\equiv \sum_a X^a
Y^a$ and $(X\times Y)^a\equiv \sum_{bc} g f^{abc} X^b Y^c$, where $f^{abc}$ are
the $su(N)$ structure constants and $g$ is the gauge coupling. In this notation
the gauge-covariant derivative in the adjoint representation is  $D_\mu
X=\p_\mu X+ A_\mu\times X$.  If one integrates out the canonically conjugate
color-electric field $\pi_i$, one gets the Coulomb-gauge Faddeev-Popov
Lagrangian density in the second-order formalism,
\begin{align}
 \L^{\rm FP} &=  \half ( D_0 A_i - \p_i A_0)^2 + (1/4) F_{i j}^2  - \partial_i\bar c\cdot D_i c + i \p_i b\cdot A_i  \,,
 \label{ko2_lag}
	\end{align} 
Next, we integrate out the $b$-field, so the gauge condition is satisfied
on-shell, and ${\bf A}$ is purely transverse,
\beq
\p_i A_i = 0.
\eeq
We separate the transverse and longitudinal parts of ${\bf \pi}$, 
\beq
\label{decomposepi}
\pi_i = \tau_i - \p_i \lambda,
\eeq
where $\p_i \tau_i = 0$.  The Faddeev-Popov action with the on-shell gauge
condition is given by
\beq
\label{FPaction}
S = \int d^{d+1}x  \left[ i \tau_i \cdot D_0 A_i 
 +  \half \tau^2   
 + (1/4) F_{i j}^2 
+ \half (\p_i\lambda)^2 + i \p_i \lambda \cdot D_i A_0 - \partial_i {\bar c}\cdot D_i c \right],
\eeq
where we have used $\int d^{d+1}x \ \tau_i \cdot \p_i A_0 = \int d^{d+1}x \
\p_i \lambda \cdot \p_0 A_i = 0$.  The time derivative appears only in
the first term, $\tau_i \cdot D_0 A_i = \tau_i \cdot \p_0 A_i +
\tau_i \cdot g A_0 \times A_i$.

\section{Time-dependent gauge transformations and their consequence for propagators} \label{sec:symmetry}

The gauge condition $\p_i A_i = 0$ does not fix time-dependent gauge
transformations $g(t)$.  Moreover the action $S$ is invariant under such gauge
transformations,
\beq
S(^g\Phi_\alpha, ^g A_0) = S(\Phi_\alpha, A_0),
\eeq
where the fields transform according to
\beqa
\label{finitegaugetransf}
\Phi_\alpha(t, {\bf x}) & \to & ^{g} \Phi_\alpha(t, {\bf x}) =  g^{-1}(t) \Phi_\alpha(t, {\bf x}) \ g(t)\\
A_0(t, {\bf x}) & \to & ^{g}A_0(t, {\bf x}) = g^{-1}(t) A_0(t, {\bf x}) \ g(t) + g^{-1}(t) \p_0 g(t),
\eeqa
and $\Phi_\alpha = t^a  \Phi_\alpha^a$ and $A_0 = t^a$ $A_0^a$.  The $t^a$
are a basis of the Lie algebra of the gauge structure group, $[t^a, t^b] = f^{abc} t^c$, and $\Phi_\alpha^a
= (A_i^a, \tau_i^a, \lambda^a, c^a, \bar c^a)$ represents all fundamental
fields besides $A_0^a$.  Under these transformations, $F_{\mu \nu}$ and $\pi$
transform gauge covariantly, $^{g}F_{\mu \nu} = g^{-1} F_{\mu \nu} g$, and
$^{g}\pi_i = g^{-1} \pi_i g$.  (In general it will be understood that $g =
g(t)$.)  A symmetry of the action implies that expectation values are invariant
under the same symmetry transformation,
\beq
\label{finitegaugeinvariance}
\left\langle O(^g\Phi_\alpha, ^g A_0) \right\rangle = \left\langle O(\Phi_\alpha, A_0) \right\rangle.
\eeq
This symmetry is generally ignored in analytic calculations, because it is
broken in usual approximation schemes.  For example, it is not a symmetry of
the tree-level theory.\footnote{Indeed the tree-level Lagrangian density in Coulomb gauge contains a time derivative in the term $\half (\p_0 A_i)^2$, and only in this term.  Under the infinitesimal time-dependent gauge transformation $\delta A_i^a = f^{abc} A_i^b \times \omega^c(t)$, this term breaks the symmetry, $\delta \half (\p_0 A_i)^2 = f^{abc}\p_0 A_i^a A_i^b \p_0 \omega^c(t) \neq 0.$    There is no other term in the tree-level Lagrangian with a time derivative to cancel this.}  However it is a powerful symmetry.

{\it Statement:} Let $\phi_1^a(x)$ and $\phi_2^a(y)$ be two fields that
transform covariantly under time-dependent gauge transformations.  Then their
propagator has a $\delta$-function singularity in time
\beq
\label{propinvargoft}
\left\langle \phi_1^a(x) \phi_2^b(y) \right\rangle = \delta^{ab} \ U_{12}( {\bf x-y }) \ \delta(x_0 - y_0).
\eeq
The proof is immediate.  The infinitesimal form of the time-dependent gauge
transformation, \eqref{finitegaugetransf}, is
\beq
\delta \phi_i(x) = \omega(x_0) \times \phi_i(x), 
\eeq
where $i = 1, 2$, and invariance under infinitesimal time-dependent gauge
transformations, \eqref{finitegaugeinvariance}, reads
\beq
\left\langle \delta [ \phi_1^a(x) \ \phi_2^b(y) ] \right\rangle
=  \left\langle [ \omega(x_0) \times \phi_1(x) ]^a \ \phi_2^b(y) + \phi_1^a(x) \ [ \omega(y_0) \times \phi_2(y) ]^b \right\rangle = 0.
\eeq
Global gauge invariance, that is, for $g = {\rm const}$, implies that
$\left\langle \phi_1^a(x) \phi_2^b(x) \right\rangle = \delta^{ab} {\cal D}(x-y)
$, and we have
\beq
f^{acb} [ \omega^c(x_0) - \omega^c(y_0) ] {\cal D}(x-y) = 0.
\eeq
This holds for all $\omega(t)$.  The general solution to this condition, which is a well defined distribution, is
\eqref{propinvargoft}, as asserted.  The proof holds for other non-trivial
representations such as the fundamental representation.  It also extends
immediately to the lattice.  Propagators whose time-dependence is given by $\delta(x_0 - y_0)$ will be
called ``instantaneous."

\section{Propagators In Coulomb gauge}

The scalar fields $A_0$ and $\lambda$ and the ghost pair $c$ and $\bar c$
appear at most quadratically in the action, \eqref{ko1_lag}, with fixed $A_i$ and $\tau_i$.  To calculate the propagators of
these fields, one may integrate out the fields $A_0$ and $\lambda$ or $c$ and
$\bar c$  by Gaussian integration, and one obtains the well-known formulas
\beqa
\label{integrateout}
\delta^{ab} D_{c \bar c}(x - y) & = & \left\langle \left(M^{-1} \right)^{ab}(x) \right\rangle \delta(x_0 - y_0) \nonumber  \\
\delta^{ab} i D_{A_0 \lambda}(x - y) & = &
\left\langle   (M^{-1})^{ab}(x) \right\rangle \delta(x_0 - y_0)  
+ i \left\langle A_{0{\rm phys}}^a(x) \ \lambda_{\rm phys}^b(y)  \right\rangle
\nonumber  \\
\delta^{ab} D_{A_0 A_0}(x - y) & = & \left\langle K^{ab}({\bf x, \ y}) \right\rangle \delta(x_0 - y_0)  + \left\langle A_{0{\rm phys}}^a(x) \ A_{0{\rm phys}}^b(y)  \right\rangle 
\nonumber \\
\delta^{ab} D_{\lambda \lambda}(x - y) & = & \left\langle \lambda_{\rm phys}^a(x) \ \lambda_{\rm phys}^b(y)  \right\rangle,
\eeqa
where
\beq
M({\bf A}) = - D_i({\bf A}) \cdot \p_i
\eeq
is the d-dimensional
Faddeev-Popov operator that depends only on the transverse dynamical field $A_i$, $K$ is the operator with
kernel
\beq
K^{ab}({\bf x, y}; y_0) = \left[ M^{-1} ( - \nabla^2 ) M^{-1} \right]^{ab}({\bf x},{\bf y}; y_0),
\eeq
and
\beqa
\label{physval}
\lambda_{\rm phys}^a(x) & \equiv & \int d^dy \ (M^{-1})^{ab}({\bf x, \ y}; \ x_0) \ \rho^b({\bf y}, x_0)    \nonumber \\
iA_{0{\rm phys}}(x) & \equiv & - \int d^dy \ K^{ab}({\bf x, \ y}; \ x_0) \ \rho^b({\bf y}, x_0),
\eeqa
are the potentials produced by the color charge density $\rho \equiv g \tau_i \times A_i$ of the dynamical gluons (and of quarks, if quarks are present).  The Faddeev-Popov operator is hermitian, $- D_i({\bf A}) \p_i = - \p_i D_i({\bf A})$, because $A_i$ is transverse, $\p_i A_i = 0$.

\section{Schwinger-Dyson equations}\label{sdses}

We wish to explore the hypothesis that there exists an asymptotic infrared limit of the DSE which is dominated by loops containing an instantaneous propagator.  Details of the derivation of the SD equations are given in
\cite{Alkofer:2009dm}.  (There is a slight change of notation.   The substitutions from \cite{Alkofer:2009dm} to the present article are $\phi \to
\lambda, \  \pi_i^T \to \tau_i,   \  A_i^T \to A_i $.)

 The time derivative appears in the action \eqref{FPaction} only once, in the canonical term $i \tau \cdot \p_0 A_i$, so the fields $\tau_i$ and $A_i$ propagate in time, and there is no instantaneous term, with factor $\delta(x_0 - y_0)$, in the propagators $D_{A_i A_j}, D_{\tau_i \tau_j}, D_{A_i \tau_j}$.  The only propagators with the instantaneous factor $\delta(x_0 - y_0)$ occur in the equations (\ref{integrateout}) for the scalar propagators.  In the DSE there are some loops that contain at least one factor of $\delta(x_0 - y_0)$, and some loops that contain none.  The DSE holds separately for each of these sets, and we shall retain only those loops that contain at least one factor of $\delta(x_0 - y_0)$. (It will turn out happily that this gives us a closed system of equations.)  Because the fourier transform of an instantaneous propagator is independent of $k_0$, 
\beq
 \int d^d {\bf x} dx_0 \ \exp[- i ( k_0 x_0 + {\bf k \cdot x}) ] \ D({\bf x}) \ \delta(x_0) =  \widetilde D({\bf k}),
\eeq
we obtain the instantaneous parts by making the substitutions
\beqa
\label{substitutions}
D_{A_0 A_0}(| {\bf k} |, k_0) & \to & D_{A_0 A_0}(| {\bf k} |)
\nonumber \\
iD_{A_0 \lambda}(| {\bf k} |, k_0) & \to & iD_{A_0 \lambda}(| {\bf k} |) = D_{c \bar c}(| {\bf k} |)
\nonumber  \\
D_{\lambda \lambda}(| {\bf k} |, k_0) & \to & 0.
\eeqa

\begin{figure}
        \centering
        \includegraphics[width=12cm]{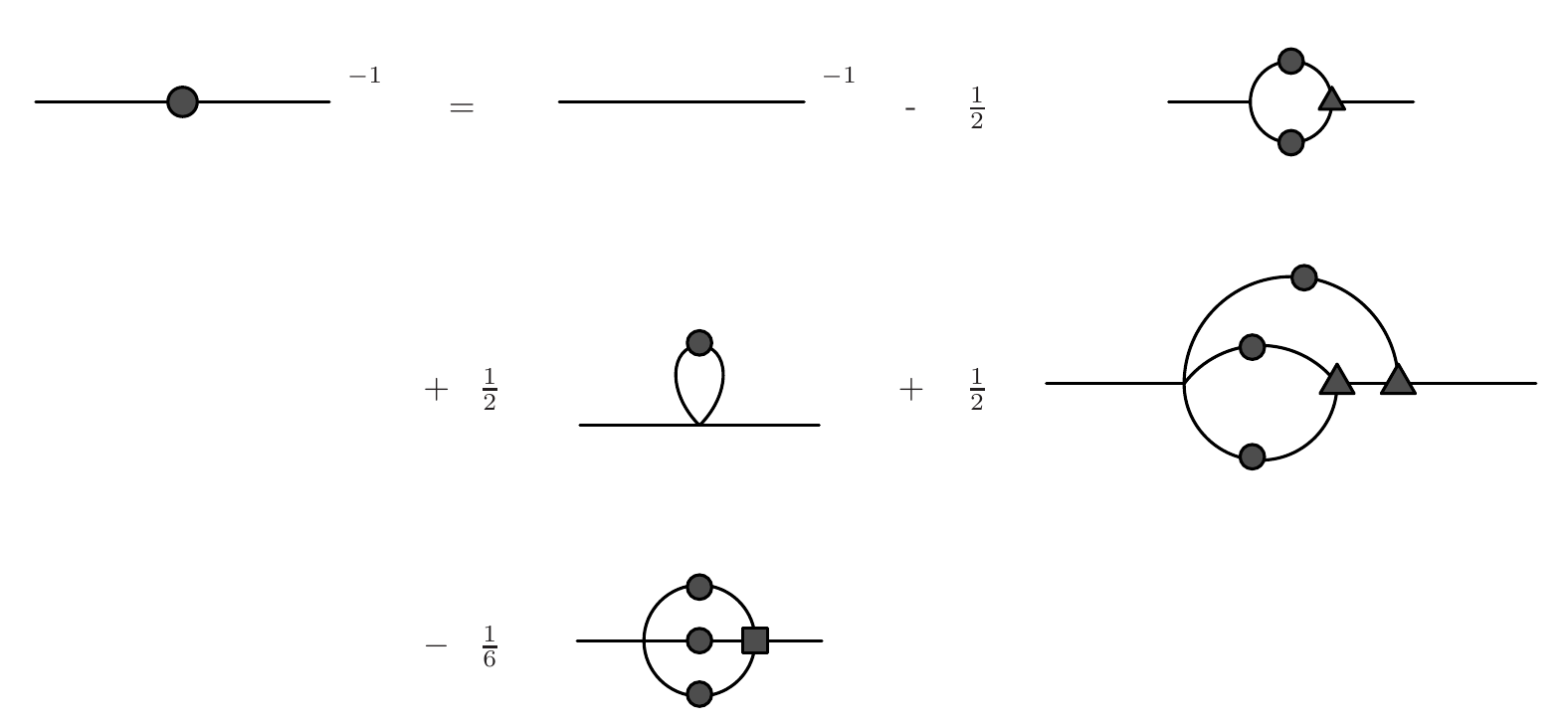}
        \caption{Graphic representation of the SD equation.  The undressed line and vertices represent tree-level quantities.  The circles, triangles and square represent dressed propagators, dressed 3-vertices, and dressed 4-vertices respectively.}
        \label{generic}
\end{figure}

\begin{figure}
        \centering
        \includegraphics[width=15cm]{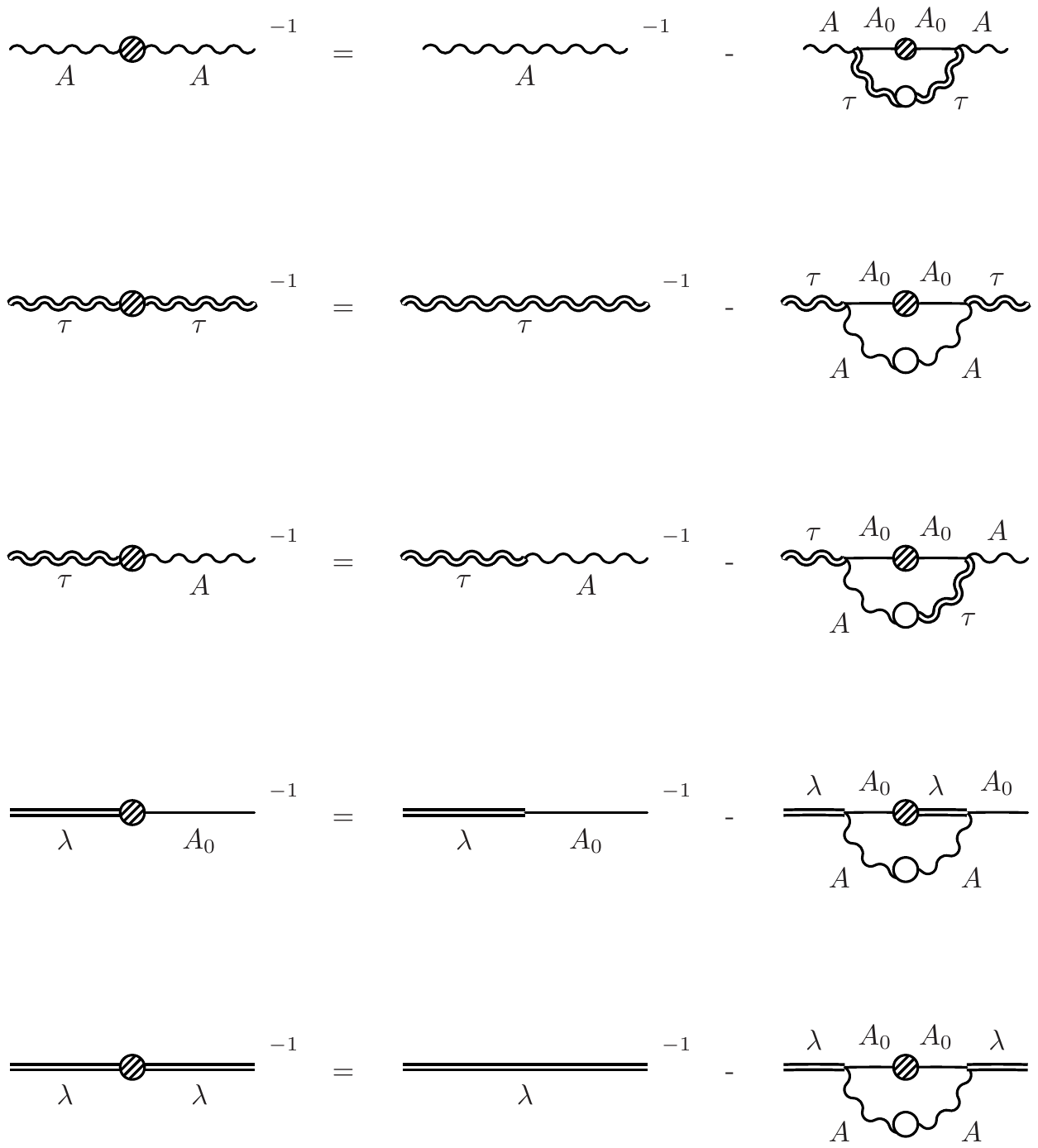}
        \caption{Diagrammatic representation of the SDE.  The shaded circle represents an instantaneous propagator and the empty circle an equal-time propagator.  The instantaneous propagators are straight lines and the equal-time propagators are wavy.}        \label{equationsFP}
\end{figure}

The SDE is represented graphically in Fig.\ \ref{generic}.  However most terms
vanish.  We discard those, and keep the remaining terms.  The tree-level terms
are retained.  The renormalization term (penguin diagram) is canceled by a mass counter-term.
Consider the other one-loop graph which is the product of two propagators.  The
possibilities are: both propagators are instantaneous, or one is instantaneous
and the other is not, or neither is.  If they are both
instantaneous, such as $\delta^2(x_0 - y_0) D_{\lambda A_0}({\bf x-y})
D_{\lambda A_0}({\bf x-y})$, there is a terrible divergence, characteristic of
the Coulomb gauge.  Fortunately these terms cancel, as we shall see shortly.
If one propagator is instantaneous, such as $V({\bf x - y}) \delta(x_0 - y_0)$, and the other, $D_N(x-y)$, is non-instantaneous, the product is instantaneous,
\beq
D_N(x-y) \ V({\bf x - y}) \delta(x_0 - y_0) = U({\bf x - y}) \delta(x_0 - y_0)
\eeq
where $U({\bf x - y}) = D_N({\bf x - y}, \ x_0-y_0 = 0) \ V({\bf x - y})$,
and gives an instantaneous contribution to $\Gamma$.  The instantaneous one-loop graphs are represented in Fig.\ \ref{equationsFP} (where the dressed 3-vertices have been
replaced by the tree-level 3-vertices, as will be discussed shortly).  If both
propagators in the loop are non-instantaneous, the result is neglected, because the product does not have a factor of
$\delta(x_0 - y_0)$, and is not instantaneous.  Now consider the two two-loop graphs in
Fig.\ref{generic}.  Both of these graphs contain a tree-level 4-vertex which
originates from the quartic $(A_i \times A_j)^2$ term in the action
\eqref{FPaction}.\footnote{There is no $(A_0\times A_i)^2$ term in the action.
It is replaced by a cubic term in $i \pi_i F_{0i}$.}  Three propagators emerge
from the tree-level 4-vertex.  Each of these propagators starts from the vector
field $A_i$ so none of them is instantaneous.  It follows that their product is not instantaneous, and their contribution may be neglected.
So far our calculations are exact.  We now make our only truncation: replace
the remaining dressed 3-vertex (in the graph in Fig.\ \ref{generic}) by the
corresponding tree-level vertex.  The result is given in Fig.\
\ref{equationsFP} and in the following equations.  This truncation has been
explored in depth, and is found to be robust numerically in both Coulomb and
Landau gauge \cite{Blum:2015lsa, Blum:2014gna, Watson:2010cn, Reinhardt:2010xm,
Huber:2009tx, Fischer:2006vf, Fischer:2009tn, Schleifenbaum:2006bq,
Schleifenbaum:2004id, Cucchieri:2004sq}.  This results from two properties of
the ghost-ghost-gluon vertex in Coulomb gauge \cite{Zwanziger:1998ez}:   (1)
The external ghost momenta factor out of the corresponding Feynman integrals.
This depresses the degree of convergence of the integrals, so (2) the vertex
does not require renormalization $\widetilde Z_1 = 1$.  These properties
severely restrict the allowed form  of the complete vertex, and investigation
did not reveal a new acceptable solution of the SD equation
\cite{Alkofer:2009dm}.  The same properties hold in the Landau gauge
\cite{Bloch:2003yu}. 

  The resulting
equations are represented graphically in Fig. \ref{equationsFP}, and,
analytically by
\beqa
\Gamma_{{\bf A}{\bf A}}(k) &  = & {\bf k}^2+ { Ng^2 \over (2\pi)^{d+1} } \int d^{d+1}p \Big[ {d-2 + (\hat{k} \cdot \hat{p})^2 \over d-1} \ D_{{\bf \tau \tau}}(p) \  D_{A_0A_0}(k-p)  
\nonumber \\
\label{dsatrans}
&& \hspace{0.7cm}+ {1 - (\hat{k} \cdot \hat{p})^2 \over d-1} \ {\bf p}^2  \ \left( \ D_{A_0 A_0}(p) \ D_{\lambda \lambda}(k-p) + \ D_{A_0 \lambda}(p) \ D_{A_0 \lambda}(k-p) + \ D_{c \bar{c}}(p) \  D_{c \bar{c}}(k-p) \right)\Big]\\
\label{dspiepie}
\Gamma_{{\bf \tau \tau}}(k) & = &  1+ { Ng^2 \over (2\pi)^{d+1} } \int d^{d+1}p \  {d-2 + (\hat{k} \cdot \hat{p})^2 \over d-1}  D_{{\bf A}{\bf A}}(p) \  D_{A_0A_0}(k-p)  \\
\label{dspiea}
\Gamma_{{\bf \tau A}}(k) & = &  -k_0 + { Ng^2 \over (2\pi)^{d+1} } \int d^{d+1}p \ {d-2 + (\hat{k} \cdot \hat{p})^2 \over d-1}\ D_{{\bf A \tau}}(p) \  D_{A_0A_0}(k-p) \\
\label{ddazeroazero}
\Gamma_{A_0A_0}(k) & = & {Ng^2 \over (2\pi)^{d+1}} \int d^{d+1}p \  {\bf k}^2[1 - (\hat{k} \cdot \hat{p})^2] \ D_{{\bf A}{\bf A}}(p) \  D_{\lambda \lambda}(p-k)  \\
\Gamma_{\lambda \lambda}(k) & = & {\bf k}^2+ {Ng^2 {\bf k}^2\over (2\pi)^{d+1}} \int d^{d+1}p \ [1 - (\hat{k} \cdot \hat{p})^2] \ D_{{\bf A}{\bf A}}(p) \  D_{A_0 A_0}(p-k) \\
\label{dsphiazero}
\Gamma_{\lambda A_0}(k) & = & i {\bf k}^2+ {Ng^2{\bf k}^2 \over (2\pi)^{d+1}} \int d^{d+1}p \ [1 - (\hat{k} \cdot \hat{p})^2] \ D_{{\bf A}{\bf A}}(p) \  D_{A_0 \lambda}(p-k) \\ 
\label{dsghost}
\Gamma_{\lambda A_0}(k) & = & i\Gamma_{\bar{c}  c}(k),
\eeqa
where $D_{c \bar c} = \Gamma_{\bar c c}^{-1}$, and $\hat{k}$ and $\hat{p}$ are
unit vectors.  Note that the vector  propagators are functions of $p$, and the
scalar propagators are functions of $k - p$.  The terms on the left-hand side
$\Gamma_{\alpha \beta}$ are the two-point functions of the quantum effective
action and are inverse to the propagators 
\beq
\Gamma_{\alpha \beta} D_{\beta \gamma} = \delta_{\alpha \gamma},
\eeq
where the indices run over all degrees of freedom.  As discussed above, terms on the right-hand side such as $D_{{\bf A} {\bf
A}}(p) D_{{\bf A} {\bf A}}(k-p)$ (and quark loops if any), where both factors
are non-instantaneous, do not contribute to these SD equations.

As follows from \eqref{substitutions}, we set $D_{\lambda \lambda} = 0$ on the right hand side of the SD equations where it appears, namely in the first term of the second line of \eqref{dsatrans}.  The remaining two terms in the second line of \eqref{dsatrans} appear to suffer from
terrible divergences.  In position space each is the product of two
instantaneous propagators.  For example the second term in that line is
$\delta^2(x_0 - y_0) D_{\lambda A_0}({\bf x-y}) D_{\lambda A_0}({\bf x-y})$.  The second and third terms contain
the divergent integral $\int dp_0 = \infty$, which is the momentum-space
manifestation of the divergent factor $\delta^2(x_0 - y_0)$.  These are the
famous energy divergences of the Coulomb gauge which cancel between the second
and third term \cite{Mohapatra:1971, Andrasi:2014nxa, Andrasi:2005xu,
Andrasi:2015pha},
\beq
\int dp_0 \ \left( D_{A_0 \lambda}({\bf p}) \ D_{A_0 \lambda}({\bf k - p}) + \ D_{c \bar{c}}({\bf p}) \  D_{c \bar{c}}({\bf k - p}) \right) = 0,
\eeq
by virtue of $D_{c \bar c} = i D_{A_0 \lambda}$, \eqref{substitutions}.  Thus all three terms in the second line in \eqref{dsatrans}
are conveniently eliminated.  (This argument is not rigorous because there remain unresolved ambiguities in the Coulomb gauge \cite{Andrasi:2014nxa}.) Each of the remaining terms in the SDE is the
product of an instantaneous propagator and an equal-time propagator, which
together give a finite instantaneous contribution.

With these results the DSE simplifies to
\beqa
\Gamma_{{\bf A}{\bf A}}({\bf k}) &  = & {\bf k}^2+ { Ng^2 \over (2\pi)^{d+1} } \int d^{d+1}p \  {d-2 + (\hat{k} \cdot \hat{p})^2 \over d-1} \ D_{{\bf \tau \tau}}(p) \  D_{A_0A_0}({\bf k - p})
\label{dsAAb} \\
\label{dspiepieb}
\Gamma_{{\bf \tau \tau}}({\bf k}) & = &  1+ { Ng^2 \over (2\pi)^{d+1} } \int d^{d+1}p \  {d-2 + (\hat{k} \cdot \hat{p})^2 \over d-1}  D_{{\bf A}{\bf A}}(p) \  D_{A_0A_0}({\bf k-p})  \\
\label{dspieb}
\Gamma_{{\bf \tau A}}(k) & = &  -k_0 + { Ng^2 \over (2\pi)^{d+1} } \int d^{d+1}p \ {d-2 + (\hat{k} \cdot \hat{p})^2 \over d-1}\ D_{{\bf A \tau}}(p) \  D_{A_0A_0}( {\bf k-p} ) \\
\label{ddazeroazerob}
\Gamma_{A_0A_0}({\bf k}) & = &  0 \\
\Gamma_{\lambda \lambda}({\bf k}) & = & {\bf k}^2+ {Ng^2 {\bf k}^2\over (2\pi)^{d+1}} \int d^{d+1}p \ [1 - (\hat{k} \cdot \hat{p})^2] D_{{\bf A}{\bf A}}(p) \  D_{A_0 A_0}({\bf k-p}) \\
\label{dsphiazerob}
\Gamma_{\lambda A_0}({\bf k}) & = & i {\bf k}^2+ {Ng^2{\bf k}^2 \over (2\pi)^{d+1}} \int d^{d+1}p \ [1 - (\hat{k} \cdot \hat{p})^2] D_{{\bf A}{\bf A}}(p) \  D_{A_0 \lambda}({\bf k-p}).
\eeqa

\section{Equal-time propagator from the loop integral}
\label{sec:eqtprops}

  Consider the loop integral for $\Gamma_{{\bf A}{\bf A}}({\bf k})$.  The only
  appearance of $p_0$ in the integrand occurs in $D_{{\bf \tau \tau}}(p_0, {\bf
  p})$, so the loop integral over $p_0$ takes the form
  \beq
\label{equaltimeprop}
\int { d p_0 \over 2 \pi}  \ D_{{\bf \tau \tau}}(p_0, {\bf p}) = D_{{\bf \tau \tau}}^{ET}({\bf p}), 
\eeq
where the right-hand side is the equal-time propagator.  Indeed it is a special
case  of the fourier transform,
  \beq
  \widetilde D_{{\bf \tau \tau}}(t, {\bf p}) = \int { d p_0 \over 2 \pi } \exp(i p_0 t) D_{\bf \tau \tau}(p_0, {\bf p}),
\eeq
 at t = 0, $D_{\bf \tau \tau}^{ET}({\bf p}) = \widetilde D_{\bf \tau \tau}(0,
 {\bf p}) $.  The remaining integration $\int d^dp$ is an integral over the
 space dimension~$d$.  The same is true for all the loop integrals.       
  
  We now show that 
  \beq
  \Gamma_{{\bf \tau A}}(k_0, {\bf k}) = - k_0.
  \eeq
  is a solution of \eqref{dspieb}.  Indeed, suppose this is true.  It gives
\begin{equation}
\label{anotherinvert}
\left(
\begin{array}{rlc}
        D_{{\bf \tau \tau}}(k_0, {\bf k}) & \ D_{{\bf \tau A}}(k_0, {\bf k})   \\
        D_{{\bf A \tau}}(k_0, {\bf k})  & \  D_{{\bf A} {\bf A}}(k_0, {\bf k})
\end{array}
\right)
= \left(
\begin{array}{rlc}
        \Gamma_{{\bf \tau \tau}}({\bf k}) \ & \ \ \ \ - k_0   \\
    k_0 \ \ \ \    & \ \ \Gamma_{{\bf A}{\bf A}}({\bf k})
\end{array}
\right)^{-1}
= { 1 \over k_0^2 
+ \Gamma_{{\bf \tau \tau}}({\bf k}) \Gamma_{{\bf A} {\bf A}}({\bf k}) }
\left(
\begin{array}{rlc}
 \Gamma_{{\bf A} {\bf A}}({\bf k}) \ & \ \ \ \ k_0   \\
        - k_0 \ \ \   & \ \ \Gamma_{{\bf \tau \tau}}({\bf k})
\end{array}
\right),
\end{equation}   
 which implies
 \beq
 \int { d p_0 \over 2 \pi}  \
 D_{{\bf A \tau}}(p_0, {\bf p}) = 0,
 \eeq
 because $D_{{\bf A \tau}}(p_0, {\bf p})$ is odd in $p_0$.  It follows that the integral in \eqref{dspieb} vanishes, which gives $\Gamma_{{\bf \tau A}}(k_0, {\bf k}) = - k_0$ so \eqref{dspieb} is satisfied.  (There may also be a non-perturbative solution which cannot be expressed as a power series in g.)  $\Box$

The SD equations now read
\beqa
\label{AAc}
\Gamma_{{\bf A}{\bf A}}({\bf k}) &  = & {\bf k}^2+ { Ng^2 \over (2\pi)^d } \int d^dp \ {d-2 + (\hat{k} \cdot \hat{p})^2 \over d-1} \ D_{{\bf \tau \tau}}^{ET}({\bf p}) \  D_{A_0A_0}({\bf k - p}) \\
\label{dspiepiec}
\Gamma_{{\bf \tau \tau}}({\bf k}) & = &  1+ { Ng^2 \over (2\pi)^d } \int d^dp \  {d-2 + (\hat{k} \cdot \hat{p})^2 \over d-1}  D_{\bf A A}^{ET}({\bf p}) \  D_{A_0A_0}({\bf k-p})  \\ 
\label{dspiec}
\Gamma_{{\bf \tau A}}(k) & = &  -k_0  \\
\label{ddazeroazeroc}
\Gamma_{A_0A_0}({\bf k}) & = &  0 \\
\label{dslambdalambdac}
\Gamma_{\lambda \lambda}({\bf k}) & = & {\bf k}^2+ {Ng^2 {\bf k}^2\over (2\pi)^d} \int d^dp \ [1 - (\hat{k} \cdot \hat{p})^2] D_{\bf A A}^{ET}({\bf p}) \  D_{A_0 A_0}({\bf k-p}) \\
\label{dsphiazeroc}
\Gamma_{\lambda A_0}({\bf k}) & = & i {\bf k}^2+ {Ng^2{\bf k}^2 \over (2\pi)^d} \int d^dp \ [1 - (\hat{k} \cdot \hat{p})^2] D_{\bf A A}^{ET}({\bf p}) \  D_{A_0 \lambda}({\bf k-p}).
\eeqa
Upon close inspection of equation \eqref{anotherinvert}, something may seem
amiss. The $k_0$ dependence in the propagators implies that these propagators
are non-instantaneous, which violates the symmetry discussed in section
\ref{sec:symmetry}. A brief calculation with a power-law ansatz is provided in
Appendix \ref{app:width_calc} to show that in the infrared limit, this symmetry
is restored, exhibiting a remarkable self consistency of the approach.

\section{Reduction to three unknowns}

The propagators and inverse propagators, $D_{rs} \Gamma_{st} = \delta_{rt}$, of the scalar fields are related by 
\begin{equation}
\label{invert}
\left(
\begin{array}{rlc}
D_{\lambda \lambda}({\bf k}) \ & \ D_{\lambda A_0}({\bf k})   \\
 D_{A_0 \lambda}({\bf k})  & \  D_{A_0 A_0}({\bf k})
\end{array}
\right)
= \left(
\begin{array}{rlc}
 \Gamma_{\lambda \lambda}({\bf k}) \ & \Gamma_{\lambda A_0}({\bf k})   \\
   \Gamma_{A_0 \lambda}({\bf k})     &  \Gamma_{A_0 A_0}({\bf k}) 
\end{array}
\right)^{-1}
=
\left(
\begin{array}{rlc}
 0 \ \ \ \ \ \ \ \ \ \ \  &  \ \ \ \ \ \ \ \ \ \ \left[  \Gamma_{A_0 \lambda}({\bf k}) \right]^{-1}   \\
  \left[ \Gamma_{\lambda A_0}({\bf k}) \right]^{-1} \ \ \   & \ \ - \Gamma_{\lambda \lambda}({\bf k}) \left[ \Gamma_{\lambda A_0}({\bf k}) \right]^{-2}
\end{array}
\right),
\end{equation}
where we have used $\Gamma_{A_0 A_0}({\bf k}) = 0$.  This gives $D_{\lambda \lambda}({\bf k}) = 0$, in accordance with 
\eqref{substitutions}.

Correspondingly for the dynamical propagators, we have \eqref{anotherinvert}, which gives for the equal-time propagators
\beq
\left(
\begin{array}{rlc}
    D_{{\bf \tau \tau}}^{ET}({\bf k}) & \ D_{{\bf \tau A}}^{ET}({\bf k})   \\
    D_{{\bf A \tau}}^{ET}({\bf k})  & \  D_{{\bf A} {\bf A}}^{ET}({\bf k})
\end{array}
\right)
= 
\int {dk_0 \over 2 \pi} \ 
{ 1 \over k_0^2 
+ \Gamma_{{\bf \tau \tau}}({\bf k}) \Gamma_{{\bf A} {\bf A}}({\bf k}) }
\left(
\begin{array}{rlc}
 \Gamma_{{\bf A} {\bf A}}({\bf k}) \ & \ \ \ \ k_0   \\
        - k_0   & \ \ \Gamma_{{\bf \tau \tau}}({\bf k})
\end{array}
\right)
\eeq
\beq
\label{equaltimeidentityFP}
\left(
\begin{array}{rlc}
D_{\tau \tau}^{ET}({\bf k}) & \ D_{\tau {\bf A}}^{ET}({\bf k})   \\
 D_{{\bf A} \tau}^{ET}({\bf k})  & \  D_{{\bf A} {\bf A}}^{ET}({\bf k})
\end{array}
\right)
= 
{ 1 \over 2 }
\left(
\begin{array}{rlc}
 (\Gamma_{{\bf A} {\bf A}} / \Gamma_{\tau \tau})^{1/2}({\bf k}) \ & \ \ \ \ 0   \\
   0   & \ \ (\Gamma_{\tau \tau} / \Gamma_{{\bf A}{\bf A}})^{1/2}({\bf k})
\end{array}
\right).
\eeq
Note that $D_{{\bf \tau A}}(k_0, {\bf k})$ is odd in $k_0$ which gives $D_{{\bf \tau  A}}^{ET}({\bf k}) = \int dk_0 D_{{\bf \tau A}}(k_0, {\bf k}) = 0$, as claimed.

From the last equation we have the simple identity,
\beq
4 D_{{\bf \tau \tau}}^{ET}({\bf k}) D_{{\bf A A}}^{ET}({\bf k}) = 1,
\eeq
which determines $D_{{\bf \tau} \tau}^{ET}$.  There remain only three
independent unknown functions $D_{\bf A A}^{ET}({\bf k}), D_{A_0 A_0}({\bf k})$
and $D_{A_0 \lambda}({\bf k})$.
We also have from \eqref{equaltimeidentityFP},
\beq
\label{eqforDAAETsquare}
\left[ D_{{\bf A A}}^{ET}({\bf k}) \right]^2 = { \Gamma_{\tau \tau}({\bf k}) \over 4 \Gamma_{{\bf A A}}({\bf k}) }.
\eeq
The last two equations give
\beq
{\Gamma_{\bf A A}({\bf k}) \over D_{\tau \tau}^{ET}({\bf k})} = { \Gamma_{\tau \tau}({\bf k}) \over D_{\bf A A}^{ET}({\bf k}) }.
\eeq

We now substitute the right hand side of the SDE for $\Gamma_{\bf A A}({\bf k})$ and $\Gamma_{\tau \tau}({\bf k})$, \eqref{AAc} and \eqref{dspiepiec}, into the last equation, which gives
\beqa
\label{ISDforDETyy}
4 {\bf k}^2 D_{\bf A A}^{ET}({\bf k})  - [ D_{\bf A A}^{ET}({\bf k}) ]^{-1} 
& = &  g^2 N \int { d^dp \over (2 \pi)^d } \ { d - 2 + (\hat p \cdot \hat k)^2 \over d-1 } \ \left( {D_{\bf A A}^{ET}({\bf p}) \over D_{\bf A A}^{ET}({\bf k}) } -  { D_{\bf A A}^{ET}({\bf k}) \over  D_{\bf A A}^{ET}({\bf p}) } \right)  \ D_{A_0 A_0}({\bf p} - {\bf k}) \\
\label{ISDforDETaa}
D_{A_0 A_0}({\bf k}) \left[ D_{c \bar c}({\bf k}) \right]^{-2} & = & {\bf k}^2+ {Ng^2 {\bf k}^2\over (2\pi)^d} \int d^dp \ [1 - (\hat{k} \cdot \hat{p})^2] D_{\bf A A}^{ET}({\bf p}) \  D_{A_0 A_0}({\bf k-p}) \\
\label{dsphiazerocz}
\left[ D_{c \bar c}({\bf k}) \right]^{-1} & = & {\bf k}^2- {Ng^2{\bf k}^2 \over (2\pi)^d} \int d^dp \ [1 - (\hat{k} \cdot \hat{p})^2] D_{\bf A A}^{ET}({\bf p}) \  D_{c \bar c}({\bf k-p}),
\eeqa
where the last two equations come from \eqref {dslambdalambdac} and \eqref{dsphiazeroc}, and we have used
\beq
\Gamma_{\lambda A_0} = i \Gamma_{\bar c c} = 
i [ D_{c \bar c} ]^{-1} = [ D_{\lambda A_0} ]^{-1} .
\eeq

Altogether there are three equations for the three propagators  $D_{\bf A
A}^{ET}({\bf k}), D_{c \bar c}({\bf k})$ and $D_{A_0 A_0}({\bf k}).$  These
three quantities are invariant under the remnant gauge symmetry $g(t)$.
Suppose the three equations are solved, so these three quantities are known.
Then one can recover a 4th quantity, $D_{\tau \tau}^{ET}({\bf k})$, from $4
D_{\tau \tau}^{ET}({\bf k}) D_{\bf A A}^{ET}({\bf k}) = 1$.  These 4 quantities are
all that appear on the right hand side of eqs. (\ref{AAc}) through
(\ref{dsphiazeroc}), from which one can recover all $\Gamma_{\alpha \beta}$ and
hence all propagators $D_{\beta \gamma}$.

\section{Gauge condition on the lattice and in the continuum}\label{horizon_maxb}

Beside imposing the Coulomb gauge condition, $\p_i A_i = 0$, we must also address the non-perturbative issue of Gribov copies \cite{GRIBOV19781, Vandersickel:2012tz, Maas:2009se, Maas:2017csm}.

A gauge choice that is accessible to numerical simulation
is implemented by minimizing (the lattice analog of) the spatial Hilbert norm,
\beq
\label{minimizingfunction}
F_A(g) \equiv \int d^{d+1}x  \ \sum_{i = 1}^d({^g}A_i^b)^2, 
\eeq
with respect to gauge transformations $g(x)$, where $A_\mu = \half i \tau^b A_\mu^b$ and ${^g}A_\mu = g^{-1}A_\mu g + g^{-1} \p_\mu g$. At a global or local minimum, the gauge condition $\p_i A_i = 0$ is satisfied, and all eigenvalues of the Faddeev-Popov operator $M(A)$ are non-negative $\lambda_n(gA) \geq 0$.  The set of continuum configurations that satisfy these conditions is designated by~$\Omega$ and is called the ``(first) Gribov region."  It is a convex region in configuration space (A-space) that is bounded in every direction.  Its boundary, $\p\Omega$, is called the ``Gribov horizon."  At large volume $V$, $\Omega$ is specified by $ H(gA) \leq (N^2-1)dV$, where the ``horizon function," $H(gA)$, is defined in \eqref{horizonfunctioneq} \cite{Zwanziger:1989mf}.  The actual lattice simulation with which we shall compare was gauge-fixed by finding one local minimum of the minimizing functional for each gauge orbit.\footnote{There are various gauge choices possible within $\Omega$.}

The set $\Lambda$ of absolute minima of the minimizing functional provides a
complete gauge fixing.  It would be nice if we could perform the (functional)
integral over~$\Lambda$, but we cannot, because we do not have an explicit
description of $\Lambda$ in the physical limit of large volume $V$, as we do
for $\Omega$.\footnote{However it is known (a) that $\Lambda$ is also a
bounded, convex region, which is contained in the Gribov region, $\Lambda
\subset \Omega$, (b) that part of the boundary of $\Lambda$ coincides with part
of the boundary of $\Omega$, and (c) that there are relative minima that are
Gribov copies inside $\Omega$.}  In this situation we make the approximation
which consists in integrating over $\Omega$ instead of $\Lambda$.  This
approximation introduces a certain ``gauge-fixing error," and the total error
of the present calculation is the compound of this gauge-fixing error with the
error introduced by the truncation of terms in the SDE.

In the limit of large volume $V$, the functional integral over the Gribov
region $\Omega$ gets concentrated on its surface $\p \Omega$,\footnote{This is
        easily understood.  The integral over a unit ball $0 \leq r \leq 1 $ in
        a space of dimension $N$ has radial measure $r^{N-1}dr$.  In the limit
of large $N \to \infty$, the radial measure gets concentrated on the surface
$\delta(1 - r)$.} and the cut-off at the Gribov horizon is replaced by
insertion of the factor $\delta[(N^2 -1)dV - H]$, which enforces the ``horizon
condition."\footnote{This can be converted to a local action, at the cost of
introducing additional bose and fermi ghosts \cite{Zwanziger:1992qr}.} In
Appendix \ref{horizoncemaxbe}, it is shown that the horizon condition $\langle
H \rangle = (N^2-1)dV$, and the maximum-$b$ condition, $\lim_{|\bf k| \to
0}b({\bf k}) = \infty$, are equivalent, where $b({\bf k}) \equiv {\bf k}^2 D_{c
\bar c}({\bf k})$ is the ghost dressing function.\footnote{Observe from \eqref{devalue_eqn} that for all
        the different values of ${\bf k}$, the corresponding eigenvalues all
        change sign at the same surface ${ \langle H(gA) \rangle = (N^2 - 1) d
        V}$.  This has been called ``all horizons are one horizon"
        \cite{Zwanziger:1992qr}.  This analytic result is consistent with
        lattice study in Coulomb gauge of Nakagawa et al \cite{Nakagawa:2010eh}
        who state ``Our result is consistent with the hypothesis in the
Gribov-Zwanziger scenario that the measure of the path integral is concentrated
on the part of the horizon where ``all horizons are one horizon."  This has
also been observed in Landau gauge \cite{Sternbeck:2005vs}.}

The maximum-$b$ condition states that the ghost propagator is of longer range than the electrostatic potential, which is the same as requiring that the ghost propagator $D_{c \bar c}({\bf k})$ be more singular than ${1 \over {\bf k}^2}$ at ${\bf k} = 0$, or equivalently that the inverse ghost propagator $\Gamma_{\bar c c}({\bf k}) = D_{c \bar c}^{-1}({\bf k})$ vanishes more rapidly than ${\bf k}^2$.  The last condition is imposed by subtracting the term of
order ${\bf k}^2$ on the right hand side of the SD equation for $\Gamma_{\bar c
c}$, so it reads 
\beq
\label{Gammacbarc} 
\Gamma_{\bar c c}( {\bf k} ) = [D_{c \bar c}({\bf k})]^{-1} =  -g^2N {\bf k}^2 \int { d^dp \over (2 \pi)^d } \left[ 1 - ( \hat p \cdot \hat k)^2 \right] D_{AA}^{ET} ({\bf p}) \left[  D_{c \bar c}({\bf p} + {\bf k}) - D_{c \bar c}({\bf p} ) \right].
\eeq
There is an overall coefficient ${\bf k}^2$, and the integrand vanishes at
${\bf k} = 0$, so the right hand side vanishes faster than ${\bf k}^2$.  It is not obvious whether the last integral is positive for all $\bf k$, as it
should be if $M( {\bf A} )$ is a positive matrix, so it is a nice check that when it is
evaluated below, $I(\alpha, \gamma)$, given in \eqref{Ialphagamma}, it is in
fact positive.

\section{Three equations for three critical exponents}

\subsection{First SD equation for critical exponents}

\begin{table}

    \begin{center}
        \begin{tabular}{|| c c c ||}
                \hline
                $D_{c \bar{c}}$ & $\Rightarrow$ & $a  k^{-\alpha}$ \\  
                \hline
                $g^2D_{\bf A A}^{ET}$ & $\Rightarrow$ & $c  k^{-\gamma}$ \\
                \hline
                $g^2 D_{A_0 A_0}$ & $\Rightarrow$ & $\bar d  k^{-\delta}$ \\
                \hline
        \end{tabular}
    \end{center}
    \caption{Critical exponents defined in the infrared asymptotic limit.}
    \label{criticalXexponents}
\end{table}


We now assume that the propagators approach an asymptotic limit at small $k = | {\bf k} |$ which is a power law, with critical exponents defined in Table \ref{criticalXexponents}.  We substitute this power-law Ansatz into \eqref{ISDforDETyy}, 
\beqa
\label{ISDforDETbis}
{ 4 k^2 c \over k^\gamma } - { g^2 k^\gamma \over c } 
= N \int { d^dp \over (2 \pi)^d } \ { d - 2 + (\hat p \cdot \hat k)^2 \over d-1 } \ \left( { \ k^\gamma \over \ p^\gamma }
 - { p^\gamma \over k^\gamma } \right) 
 \ {\bar d \over |\vec p - {\bf k}|^\delta }.
\eeqa 

For this to yield a bona fide solution, the loop integral must converge.  There is a singularity due to the color-Coulomb potential, $1/|{\bf p} - {\bf k}|^\delta$.  However the factor ${ \ k^\gamma \over \ p^\gamma }
 - { p^\gamma \over k^\gamma }$ vanishes at ${\bf p} = {\bf k}$ like $({\bf p} - {\bf k})^2$, so the integral converges at ${\bf p} = {\bf k}$ provided $\delta < d + 2$. There is a singularity at $p = 0$ due to the terms $p^\gamma$ and $p^{-\gamma}$, so the integral does not converge at $p = 0$ unless $d > |\gamma|$.  The loop integral must also converge at high $p$ to assure that the infrared dynamics decouples from the other degrees of freedom.  Suppose $\gamma$  is positive $\gamma > 0$.  In this case the highest power of $p$ in the last integrand comes from the power $p^\gamma$ in the second term in the parenthesis, and the integral will not converge at high $p$ unless $d + \gamma < \delta$.  Now suppose instead that $\gamma$ is negative, $\gamma < 0$.  In this case the highest power of $p$ comes from the first term in parenthesis $1 / p^\gamma$, and the loop integral will not converge unless $d - \gamma < \delta$, and we have established,
\beq
\label{confinementbound}
d < d +  |\gamma| < \delta < d + 2.
\eeq      

By power-counting one sees that the right-hand side of \eqref{ISDforDETbis} is proportional to $k^{d - \delta}$.  The inequalites just obtained imply that in the infrared asymptotic limit, $k  \to 0$, the right-hand side is dominant over each term on the left-hand side.  Indeed it dominates the first term in this limit provided $\delta - d > \gamma - 2$, that is, if $\delta > d + \gamma -2$, which holds by virtue of \eqref{confinementbound}.  Likewise it dominates the second term on the left provided $\delta - d > - \gamma$, that is, if $\delta > d - \gamma$, which is also true.  Therefore in the infrared asymptotic limit only the right-hand side survives, and the first SDE reads,
\beqa
\label{ISDforDET3}
0 = \int { d^dp \over (2 \pi)^d } \ { d - 2 + (\hat p \cdot \hat k)^2 \over d-1 } \ \left( { \ k^\gamma \over \ p^\gamma }
 - { p^\gamma \over k^\gamma } \right) 
 \ {\bar d \over |{\bf p} - {\bf k}|^\delta }.
\eeqa

The inequality $\delta > d$, just derived, is none other than the condition in space dimension $d$, for the color-Coulomb potential to be confining, $\lim_{r \to \infty}V_C(r) = \infty$, for we have\footnote{A linearly rising color-Coulomb potential, which is favored by lattice calculations in Coulomb gauge \cite{Cooper:2015sza}, corresponds to $\delta = d + 1$.}
\beq
V_C(r) = \int d^d p \ (2\pi)^{-d} \exp(i{\bf p \cdot x}) \ D_{A_0 A_0}({\bf p})  
\sim  \int d^d p \  \exp(i{\bf p \cdot x})  /p^{\delta} \sim r^{\delta - d}.
\eeq
Thus {\it \eqref{ISDforDET3} is a sufficient condition for the color-Coulomb potential to be confining.}

\subsection{Second SD equation for critical exponents}

Insertion of the power laws into (\ref{Gammacbarc}) yields\beq
\label{Gammacbarcab} 
\Gamma_{\bar c c}( {\bf k} ) = [D_{c \bar c}( {\bf k} )]^{-1} \Rightarrow  {k^\alpha \over a} 
= - N {\bf k}^2 \int { d^dp \over (2 \pi)^d } \left[ 1 - ( \hat p \cdot \hat k)^2 \right] 
{ c \over | {\bf p} |^\gamma } \ \left( { a \over |{\bf p} - {\bf k} |^\alpha } - { a \over | {\bf p} |^\alpha }  \right).
\eeq
By counting powers of $k$ and $p$ on the left and right hand sides, we obtain $\alpha = 2 + d - \gamma - \alpha$, which gives the ``sum rule"
\beq
\label{sumruleFP}
2 \alpha + \gamma =  d + 2.
\eeq
 Moreover the right hand side has a coefficient ${\bf k}^2$, and an integrand that vanishes with ${\bf k}$, so the right hand side vanishes with $k$ more rapidly than ${\bf k}^2$.  We conclude that $\alpha > 2,$ so the ghost propagator is more singular than the free propagator.\footnote{The case $d = 2$ is singular.  We deal with this by continuing in dimension to $d > 2$, and for the case $d = 2$, we take the limit $d \to 2$ at the end of the calculation.}  This was imposed by the horizon condition.  We require that the loop integral (\ref{Gammacbarcab}) converges at high $p$.  This yields the inequality $d < \gamma + \alpha + 2$.  Indeed the subtraction term cancels the leading term in $1/p$ at high $p$, and the next power is killed by angular integration, so the subtraction term increases the power of $1/p$ by~2.  We substitute
\beq
\label{sumrule}
\gamma = d + 2 - 2\alpha
\eeq into the last inequality and obtain $\alpha < 4$, and thus
\beq
\label{alphabound}
2 < \alpha < 4.
\eeq

\subsection{Third SD equation for critical exponents}

Upon insertion of the power Ansatz into \eqref{ISDforDETaa} we obtain
\beq
\label{GammalambdalambdaFPbis}
 \Gamma_{\lambda \lambda}(k) \Rightarrow { \bar d \ k^{2 \alpha}  \over a^2 \ k^\delta  }
= k^2 +  N  \int { d^dp \over (2 \pi)^d } \ { p^2 k^2 - (p \cdot k)^2 \over p^2 } { c \over | {\bf p} |^\gamma } \ { \bar d  \over |{\bf p} - {\bf k} |^\delta }.
\eeq
The integral converges provided $d - \gamma - \delta < 0$, which agrees with the confinement bound, \eqref{confinementbound}.  In this case the tree-level term is negligible compared to the loop term in the infrared asymptotic limit, and this ISD simplifies to
\beq
\label{GammalambdalambdaFPters}
 { \ k^{d + 2 - \gamma - \delta}  \over a^2  }
= N  \int { d^dp \over (2 \pi)^d } \ { p^2 k^2 - (p \cdot k)^2 \over p^2 } { c \over | {\bf p} |^\gamma } \ { 1  \over |{\bf p} - {\bf k} |^\delta }.
\eeq

\section{Determination of the infrared critical exponents}

From \eqref{Gammacbarcab} we obtain
\beq
{1 \over a^2 c } = I(\alpha, \gamma).
\eeq
where
\beq
\label{defineI}
I(\alpha, \gamma) \equiv - N | {\bf k} |^{\alpha + \gamma - d} \int { d^dp \over (2 \pi)^d } \left[ 1 - (\hat p \cdot \hat k)^2 \right] { 1 \over | {\bf p} |^\gamma } \ \left( {1 \over |{\bf p} + {\bf k} |^\alpha } - { 1 \over | {\bf p} |^\alpha }  \right).
\eeq
Likewise from \eqref{GammalambdalambdaFPters} we obtain
\beq
{1 \over a^2 c } = L(\delta, \gamma).
\eeq
where
\beq
\label{defineL}
L(\delta, \gamma) \equiv N |{\bf k}|^{\delta + \gamma - d} \int { d^dp \over (2 \pi)^d } \left[ 1 - (\hat p \cdot \hat k)^2 \right] { 1 \over | {\bf p} |^\gamma } \ { 1 \over |{\bf p} + {\bf k} |^\delta },
\eeq
which gives
\beq
I(\alpha,\gamma) = L(\delta, \gamma).
\eeq

One has
\beq
\label{Lofdeltaandgamma}
L(\delta, \gamma) = { N (d-1) \over 2 (4\pi)^{d/2} }{ \Gamma[(\delta + \gamma - d)/2] \ \Gamma[(d+2 - \delta)/2] \ \Gamma[(d - \gamma)/2] \over \Gamma[(\gamma + 2)/2] \ \Gamma[\delta/2]  \ \Gamma[(2d + 2 -\delta - \gamma)/2]  },
\eeq

The integral $I(\alpha, \gamma)$ was  evaluated in (A.17) of \cite{Zwanziger:2003cf}, with $I(\alpha) = I_G(\alpha_G)$ and $\alpha = 2 + 2
\alpha_G$, with the result \footnote{This result is the same as ignoring the subtraction term in
(\ref{defineI}) and continuing $L(\delta, \gamma)$ analytically from $\delta$
to $\alpha$, at fixed $\gamma$.}
\beq
\label{Ialphagamma}
I(\alpha, \gamma) = - L(\alpha, \gamma) =  - { N (d-1) \over 2 (4\pi)^{d/2} }{ \Gamma[(\alpha + \gamma - d)/2] \ \Gamma[(d+2 - \alpha)/2] \ \Gamma[(d - \gamma)/2] \over \Gamma[(\gamma + 2)/2] \ \Gamma[\alpha/2]  \ \Gamma[(2d + 2 -\alpha - \gamma)/2]  }.
\eeq
[$I(\alpha, \gamma)$ is positive because $\alpha + \gamma - d = 2 - \alpha < 0$, provided that $\alpha < 4$, which holds by \eqref{alphabound}.]  We have
\beq
- L(\alpha, \gamma) = L(\delta, \gamma).
\eeq
which gives
\beq
{ \Gamma[(\delta + \gamma - d)/2] \ \Gamma[(d+2 - \delta)/2]  \over \ \Gamma[\delta/2]  \ \Gamma[(2d + 2 -\delta - \gamma)/2]  } = - { \Gamma[(\alpha + \gamma - d)/2] \ \Gamma[(d+2 - \alpha)/2] \ \over \ \Gamma[\alpha/2]  \ \Gamma[(2d + 2 -\alpha - \gamma)/2]  }.
\eeq
 
 A third relation between the critical exponents is provided by \eqref{ISDforDET3}.  An obvious solution to that equation is provided by
\beq
\gamma = 0.
\eeq
We have searched diligently for another solution, but we have not found any.
This solution and the sum rule (\ref{sumrule}) then give
\beq
\alpha = \half ( d + 2),
\eeq 
and we finally obtain the following condition, expressed with a new parameter, $\theta \equiv \delta - d - 1$, 
\beqa
\label{deltaequation}
E(\theta, d) = F(d), 
\eeqa
where
\beqa
E(\theta, d) \equiv {\pi \over \cos \left({\pi \theta \over 2}\right)  \Gamma\left({d + 1 + \theta \over 2} \right) \Gamma\left({d + 1 - \theta \over 2} \right) }; \; \; \; \; F(d) \equiv  
 - {\Gamma\left({2-d \over 4} \right) \over \Gamma\left({3d + 2 \over 4} \right) }.           
 \eeqa
We have used $\Gamma(x) \Gamma(1-x) = \pi / \sin(\pi x)$.  The change of
variable from $\delta$ to $\theta$ is convenient because $E(\theta, d)$ is even
in $\theta$, $E(\theta, d) = E(-\theta, d)$.  

\begin{figure}
    \centering
    \includegraphics[width=10cm]{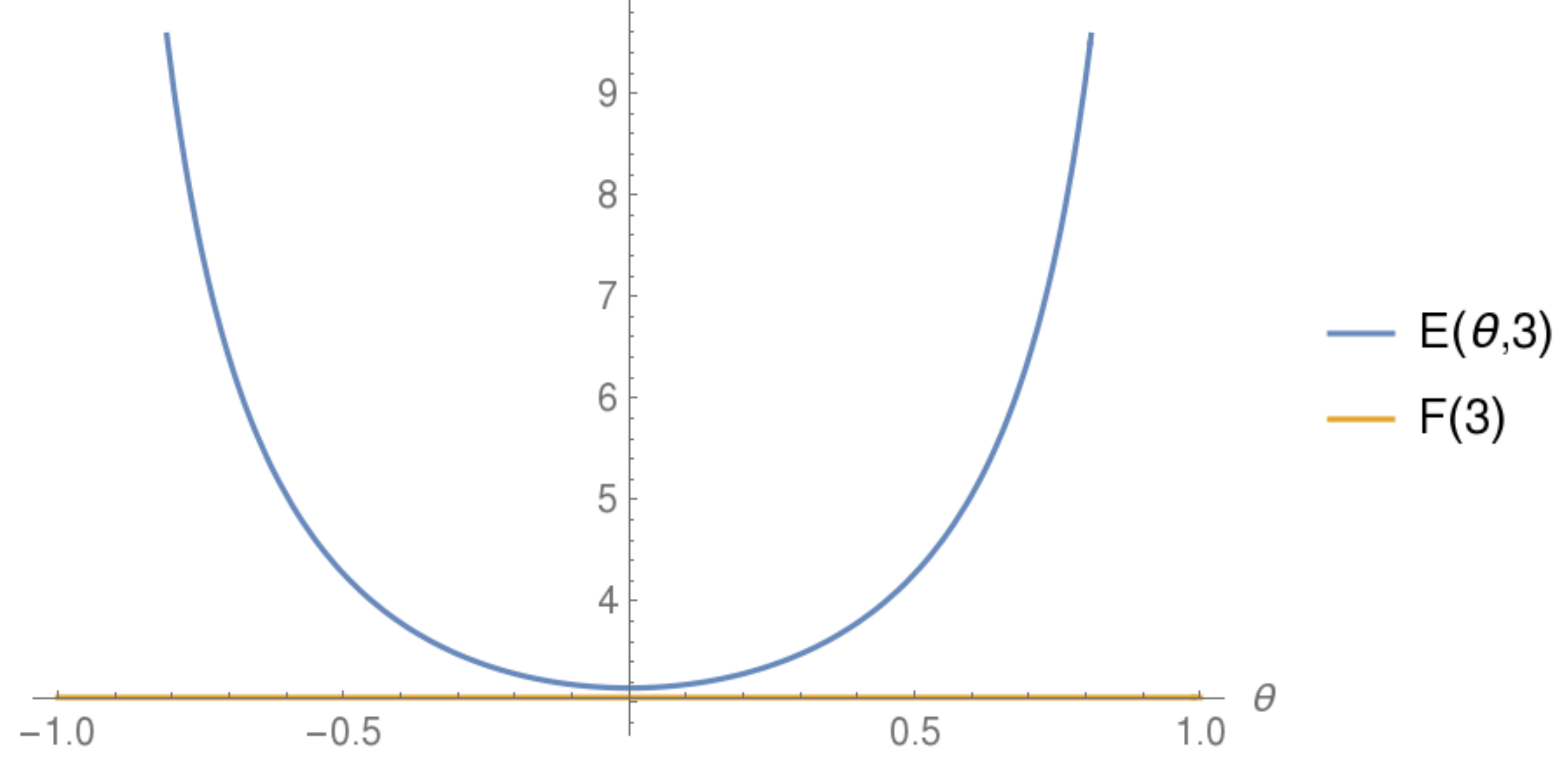}
    \caption{A near miss at space dimension $d=3$.  As $d$ decreases below $d_c$, the curved line intersects the straight horizontal line twice. }
    \label{nearmiss}
\end{figure}

For a given space dimension~$d$, let $\theta(d)$ be a solution to \eqref{deltaequation}, then the critical exponent of the color-Coulomb potential $\delta$ is recovered from
\beq
\delta(d) = d + 1 + \theta(d).
\eeq
The function $E(\theta, d)$ is finite and positive for $\theta$ in the interval
$-1 < \theta  < 1$ (which corresponds to $d < \delta < d + 2$), and is
divergent at the end-points, $\theta = \pm 1$, where $\cos( \pi \theta /2)$,
which is in the denominator, vanishes, $\cos( \pi \theta /2) = 0$.  Since
$E(\theta, d)$ is even in $\theta$, if $\theta(d)$ is a solution to
\eqref{deltaequation}, then so is $ - \theta(d)$, and the solutions to
(\ref{deltaequation}) form two branches $\theta_+(d)$ and $\theta_-(d) = -
\theta_+(d)$, as shown in Fig. \ref{nearmiss}.

\begin{figure}
    \centering
    \includegraphics[width=12cm]{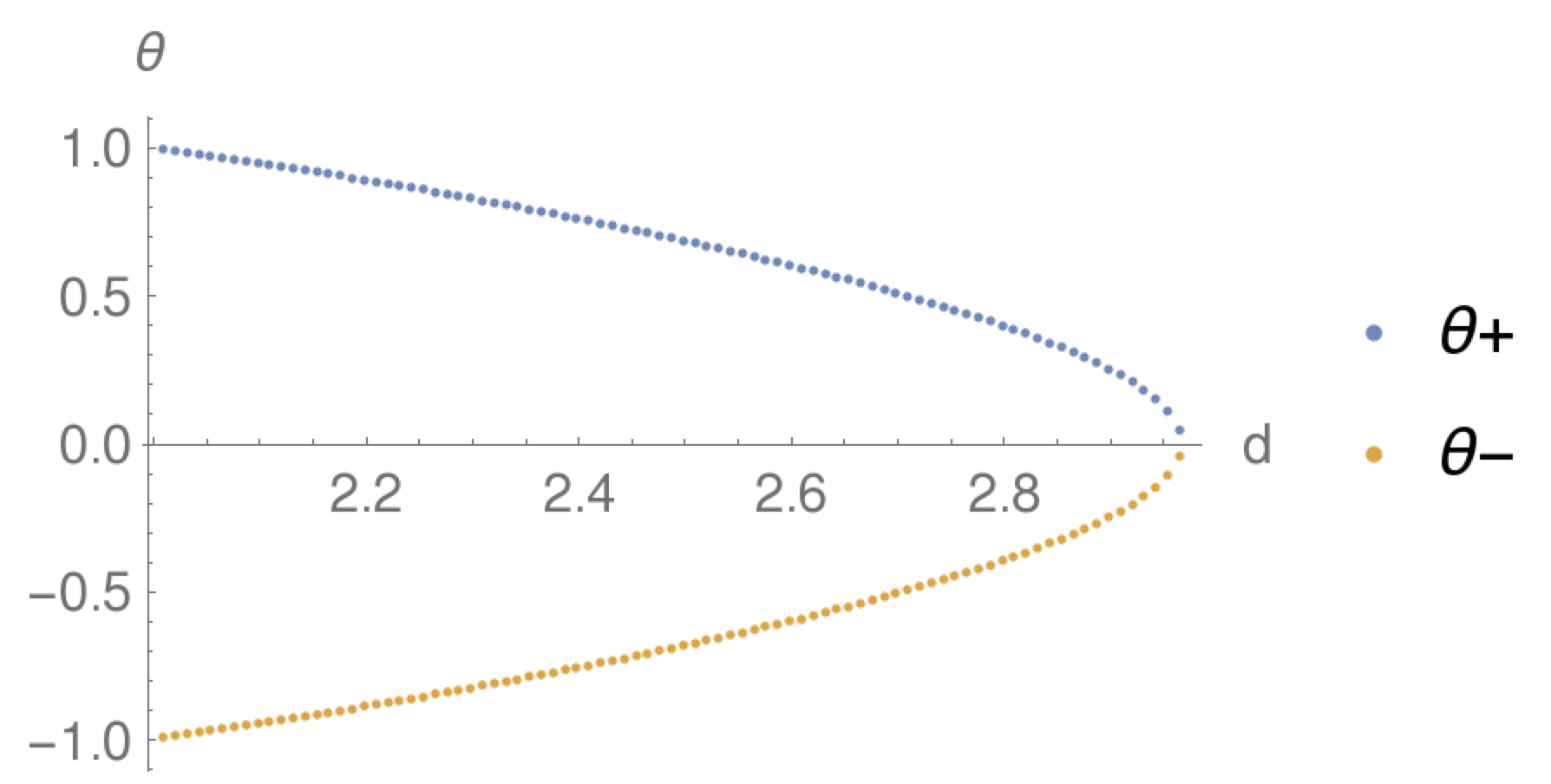}
    \caption{Plot of the two solutions $\theta_\pm(d)$.  The critical
    exponent of the color-Coulomb potential is given by $\delta = d + 1 +
    \theta_\pm(d)$.}
    \label{thetaplot}
\end{figure}

We are interested in integer space dimensions $d = 2$ and $d = 3$.  However it is helpful to take $d$ to be a continuous variable in the interval $ 2 < d < 3$.  The function $F(d)$ diverges in the limit $d \to 2$ which tells us that $\theta(2) = \pm 1$, for we have just seen that $E(\theta, \ d)$ is divergent at $\theta = \pm 1$.\footnote{The reader who wishes to avoid equating infinite quantities may prefer to solve the equation $[E(\theta, d)]^{-1} = [ F(d)]^{-1}$.}  These values of $\theta$ correspond to $\delta(d = 2) = d + 1 \pm 1 = 3 \pm 1$, so for $d = 2$, there are two solutions $\delta_-(2) = 2$ and $\delta_+(2) = 4$.  One sees in Fig.\  \ref{thetaplot} that, as $d$ increases from $d = 2$, the two branches, $\theta_+(d)$ and $\theta_-(d)$ approach each other monotonically, and at a critical dimension, 
\beq
d_c = 2.9677... \ \ ,
\eeq
they merge at $\theta = 0$, which corresponds to $\delta(\theta = 0) = d_c + 1$ \cite{Alkofer:2009dm}.  

With the problem as stated, there is no solution for space dimension above the critical value $d > d_c$.  However 2.9677 is tantalizingly close to 3, and it may be that there is a solution for $d = 3$ (a) if the gauge-fixing error noted above were corrected, (b) if the truncation error were corrected, or (c) if perhaps the true asymptotic behavior isn't a pure power-law, but rather dominated by a power with multiplicative log corrections. We shall suppose one of these possibilities is in effect, and we consider that it is an approximation to replace the physical value $d = 3$ by $d = d_c$, which is a difference of about 1\%.  For $d = 3$ and $\theta = 0$, we have
\beqa
E(\theta = 0, \ d = 3) & = & \pi = 3.14159265...
\nonumber \\
 F(d = 3) & = & 64/21 = 3.04761905... \ \ .
\eeqa  
Instead of an equality, there is a difference of about 3\%.

The color-Coulomb potential,
\beq
V_{\rm coul}(r) \sim \int d^d k \exp(i k \cdot x) {c \over k^\delta} \sim r^{\delta - d},
\eeq
is linear at large $r$ for $\delta = d + 1$, and super-linear for $\delta > d + 1$.  As space dimension $d$ increases from $d = 2$ to $d = d_c$, on the upper branch $\theta_+(d)$ decreases from $\theta_+(2) = 1$ to $\theta_+(d_c) = 0$ and is everywhere positive, $\theta_+(d) \geq 0$.  It follows that $\delta(d)$, satisfies $\delta(d) = d + 1 + \theta_+(d) \geq d + 1$.  We conclude that on the upper branch the color-Coulomb potentional $V_{\rm coul}(R)$ is everywhere  superlinear, except at $d = d_c$ where it is linear.  The lower branch is everywhere sublinear.  $V_{\rm coul}(R)$ is linear at the critical dimension $d_c$ where $\delta(d_c) = d_c + 1$.  Thus the upper branch accords with the exact theorem \cite{Zwanziger:2002sh} which asserts that the color-Coulomb potential $V_{\rm coul}(R)$ is bounded below by the gauge-invariant Wilson potential $V_W(R)$ that is linear at large $R$
\beq
V_{\rm coul}(R) \geq V_W(R) = \sigma R,
\eeq
 whereas the lower branch does not.  (Both branches accord with the exact bound $\delta > 2 \alpha -2 = d $ \cite{Cooper:2015sza}.)

The most natural choice between the two branches is to take the upper branch to be the physical solution because it satisfies the last inequality.  In $d = 2$ space dimensions, the calculation reported here yields, $\delta(d = 2) = 4$, which corresponds to a color-Coulomb potential that rises like $r^2$ at large $r$.  This is an unexpectedly steep rise.  One might speculate that the physical solution is a superposition or mixture of the two branches, so that the physical solution corresponds to a value of $\theta$ that lies between the two branches.  If so, then $\theta$ must also satisfy $\theta \geq 0$, corresponding to $\delta \geq d + 1$, to be consistent with the last inequality, and because linear rise corresponds to $\delta = d + 1$.  Finally we note that linear rise is energetically favorable compared to superlinear.  A lattice calculation in two space dimensions (if it is not already in the literature) would throw some light on this matter.

In any case the comparison we shall make with lattice gauge theory is with critical dimension $d_c = 2.9677...$ , which corresponds uniquely to linear rise, $\delta = d + 1$.

\section{Conclusion}
\label{conclusion}

\subsection{Comparison with Lattice Gauge Theory}

Lattice calculations have been reported for SU(2) \cite{Langfeld:2004qs,
Burgio:2012bk} and SU(3) \cite{Nakagawa:2009zf}.  In Table \ref{comparison} we
compare our results with Langfeld and Moyaerts \cite{Langfeld:2004qs}.  In the
first column of Table \ref{comparison} are the critical exponents, $\alpha,
\gamma$, and $\delta$, of the propagators of the ghost, the spatial gluon, and
the temporal gluon respectively, that are defined in Table
\ref{criticalXexponents}.  In the second column, the values of these exponents
found in the present article are expressed in terms of the dimension of space
$d$.  In the third column, the critical exponents defined in the present
article are expressed, for the reader's convenience, in terms of the parameters
defined by \cite{Langfeld:2004qs}.   ($\delta_{LM}$ is the infrared exponent
designated $\delta$ in \cite{Langfeld:2004qs}.)  The 4th column gives the
numerical values of the critical exponents for the critical dimension found
above, $d_c = 2.9677...\ \ .$\footnote{The numbers in the 4th column are close
to the fractions $5/2, 0, 4.$ which correspond to $d=3$.}. The final column is
result of the numerical simulation \cite{Langfeld:2004qs}.  These authors do
not give a numerical value for the infrared exponent of the transverse
equal-time gluon propagator, $\gamma$, but state ``At small momentum, the
propagator becomes roughly momentum-independent and seems to approach a
constant in the IR limit $| {\bf p} | \to 0$.''  This is consistent with our
result, which gives for this infrared exponent, $\gamma = 0$.  If it is not
accidental, the agreement between the 4th and the 5th column is remarkable for
the accuracy of both the lattice simulation and the SD equation.  

\begin{table}[!htbp]
\centering
\caption{Comparison of Schwinger-Dyson Equations to Lattice}
\begin{tabular}{*9c}
\hline
\multicolumn{3}{c}{Critical Exponents} &  & LM Notation & & SDE at $d_c$ &  & Lattice Calculation\\ 
\hline
 $\alpha$ & = & $(d+2)/2$ & = & $2 \kappa + 2$ & = & 2.4839 & $\approx$ & 2.490(10) \\ 
 $\gamma$ & = & 0 &  &  & = & 0 & $\approx$ & 0 \\ 
 $\delta$ & = & $d+1 + \theta_\pm(d) $ & = & $2 \delta_{LM}+ 2$ & = & 3.9677 & $\approx$ & 4.10(10) \\
\hline
\end{tabular}
\label{comparison}
\end{table}

\subsection{Features of Gluodynamics in the Asymptotic Infrared Limit}

We summarize the basic features of gluodynamics in the asymptotic infrared limit, under the assumption that the agreement between the SDE and the lattice gauge calculation is not accidental. 
\begin{itemize}
\item 1. The dynamics occurs in
        a single time slice.  More precisely, the ghost and temporal gluon
        propagators, $D_{c \bar c}$ and $D_{A_0 A_0}$ are both instantaneous,
        that is, proportional to $\delta(x_0 - y_0)$, and the spatial gluon
        propagator is taken at equal time, $D_{\bf A A}^{ET}({\bf x - y}) =
        D_{\bf A A}(t, {\bf x}; \ t, {\bf y})$.  This is due to the fact that
        in the Coulomb gauge, Gauss's law, $D_i \pi_i = \rho_{\rm quark}$, is a
        constraint that is satisfied as an equation of motion.

\item 2. In the asymptotic infrared limit, these propagators are fit by power
        laws with critical exponents whose values are given in the table.\\
        (a) Compared to the tree-level propagator $ 1 / | {\bf k } |^2$, the
        ghost propagator is moderately long range.\\ (b) The color-Coulomb
        propagator is long-range, corresponding to a linear rise in $r$, or
        close to it.\\  (c) The infrared limit of the equal-time spatial gluon
        propagator has critical exponent 0 or close to 0.\footnote{(a) and (b)
                are consistent with the Gribov confinement scenario. (c), on
                the other hand, is quite different. In the Gribov confinement
                scenario, the gluon propagator vanishes at $k=0$, which would
                automatically remove the gluon from the physical spectrum by
                having an unphysical spectral density. A non-vanishing, but
        constant propagator is, however, well supported in the lattice
literature, despite the challenges of extracting infrared behavior on a finite
lattice.}
  
\item 3.  The horizon condition $\langle H(gA) \rangle = (N^2 -1) d V$, and the divergence of the ghost dressing function, $\lim_{|{\bf k} \to 0|}{\bf k}^2 D_{c \bar c}({\bf k}) = \infty$, are identical gauge conditions.  This is shown in Apppendix \ref{horizoncemaxbe}, and applied in sect.~\ref{horizon_maxb} where the gauge condition is imposed by subtracting the ${\bf k}^2$ term in the SDE.

\item 4.  There is a shadow cast on these considerations because we have found no solution to the SDE at space dimension $d = 3$, but only close to it, at $d = d_c = 2.9677... \ \ $.  We must figure out what mechanism, if any, acts so there is a solution at $d = 3$.  A small effect in the right direction would be sufficient.  This could be provided by a dressed vertex replacing a tree-level vertex.

\item 5.  In Appendix \ref{CgaugeCoul}, the contribution of gluon propagators to the Wilson loop $W = N^{-1} {\rm Tr} P \exp (\oint ig t^b A_\mu^b dx_\mu)$ is calculated.  It is found that the spatial gluon propagator $D_{{\bf A}_i {\bf A}_j}$ does not contribute at all.  Only the instantaneous temporal gluon propagator contributes, which moreover exponentiates, \eqref{valueofWilson}.  Consequently the calculation of the contribution of the gluon propagator to the path-ordered exponential is particularly simple in Coulomb gauge as compared to Lorentz-covariant gauges.  This may be true for other expectation values.


\end{itemize}

\section{Acknowledgements}

Daniel Zwanziger is grateful to Richard Brandt for many stimulating conversations.

\appendix

\section{Equivalence of horizon condition and the maximum b-condition}\label{horizoncemaxbe}
 
     The boundary of the (first) Gribov region $\p \Omega$ is a set of transverse configurations $\p_i A_i = 0$ that satisfy the `horizon condition'  \cite{Vandersickel:2012tz},
\beq
    \label{horizon_condition}
    \left< H(gA) \right> = (N^2-1) d V,
\eeq
where $V = L^d$ is the spatial volume, the horizon function is defined by
\beq
\label{horizonfunctioneq}
H(gA) \equiv \int d^dx d^dy \ D_{i {\bf x}}^{ab} \ D_{i {\bf y}}^{ac} \ (M^{-1})_{{\bf x y}}^{bc},
\eeq
and $M^{ab}(gA)  = - D_i^{ab}(gA) \p_i$ is the Faddeev-Popov (FP) operator.  The functional integral over the Gribov region $\Omega$ is equivalent to the functional integral over its boundary,\footnote{See footnote above.} the Gribov horizon $\p\Omega$, and may be evaluated by insertion of $\delta[H - (N^2-1) d V]$ \cite{Zwanziger:2001kw}.  

We shall make use of two recent results \cite{Cooper:2015sza}:
\begin{enumerate} 
\item The eigenvalues of the Faddeev-Popov operator $M^{ab}(gA)$ in the asymptotic infrared region ${\bf k} \to 0$  are given by
\beq
    \label{evalue_eqn}
    \lambda_{|{\bf k}|}(gA) = {\bf k}^2 \left( 1 - { H(gA) \over (N^2 - 1) d V} +  j_{|{\bf k}|}(gA) \right),
\eeq
where $ \lambda_{|{\bf k}|}(gA)$ is the eigenvalue that emerges from  $\lambda_{|{\bf k}|}(0) = {\bf k}^2$ at $A = 0$,  and $j_{|{\bf k}|}(gA)$ vanishes as ${\bf k}$ tends to $0$,
\beq
\lim_{{\bf k} \to 0} j_{|{\bf k}|}(gA) = 0.
\eeq
In this limit we obtain
\beq
    \label{devalue_eqn}
    \lambda_{|{\bf k}|}(gA) = {\bf k}^2 \left( 1 - { H(gA) \over (N^2 - 1) d V} ) \right).
\eeq
\item In the asymptotic infrared region, the ghost propagator in an external gauge potential $\mG({\bf k}; gA) = \langle \vk |\mM^{-1}(gA) | \vk \rangle$ is given by
\beq
\label{recentresult2}
\mG^{ab}({\bf k}, gA) = { \delta^{ab} \over \lambda_{|{\bf k}|}(gA)}.
\eeq 
\end{enumerate}

The horizon function is an instance of a bulk or extensive quantity in thermodynamics which, as explained in \cite{Vandersickel:2012tz}, eq. (2.193), may be written as the integral of a density.  Typically the mean and variance of a bulk quantity satisfy,
\beq
\langle H(gA) \rangle = O(V); \hspace{2cm}H^2 - \langle H \rangle^2 = O(V).
\eeq
We assume that the horizon function behaves this way, and we easily find
\beq
H(gA) = \left\langle H(gA) \right\rangle + O(1).
\eeq
Fluctuations are down by order 1/V compared to the mean, $\langle H \rangle$.  Thus in the infinite-volume limit, $V \to \infty$, the horizon function $H(gA)$ may be replaced by its expectation-value $\langle H(gA) \rangle$, and the formula for the eigenvalues simplifies to    
\beq
    \label{evalue_eqna}
    \lambda_{|{\bf k}|}(gA) = {\bf k}^2 \left( 1 - { \langle H(gA) \rangle \over (N^2 - 1) d V} +  j_{|{\bf k}|}(gA) \right).
\eeq
We observe that when the horizon condition, \eqref{horizon_condition}, is satisfied, the term of order ${\bf k}^2$ is precisely killed, so
\beq
  \label{evalue_eqnb}
    \lambda_{|{\bf k}|}(gA) = {\bf k}^2  j_{|{\bf k}|}(gA),
\eeq
which gives
\beq
\delta^{ab} \ b({\bf k}) \equiv \delta^{ab} \ {\bf k}^2 G({\bf k}) = {\bf k}^2  \left\langle\mG^{ab}({\bf k}, gA) \right\rangle = \delta^{ab} \left\langle {1 \over  j_{|{\bf k}|}(gA)} \right\rangle.
\eeq
Thus, if the horizon condition is satisfied, the ghost dressing function is divergent at ${\bf k} = 0$,
\beq
b({\bf k} = 0) = \infty.
\eeq
We call this the maximum-$b$ condition.  The converse is also true: if the maximum-b condition is satisfied, then the horizon condition holds.  We conclude that, in the $\Omega$-theory, the horizon condition and the maximum $b$-condition $b(0) = \infty$, are equivalent.  Thus it is justified to subtract the ${\bf k}^2$ term in the equation for the inverse ghost propagator $G^{-1}({\bf k})$.

\section{Contribution of Coulomb gauge propagators to Wilson loop} \label{CgaugeCoul}

We calculate the contribution of propagators to the Wilson loop, a gauge-invariant quantity.  

Consider the expansion of the Wilson loop
\beq
W = N^{-1} {\rm Tr} \ P \exp\left( i g \oint  t^a A^a_\mu dx_\mu \right) = N^{-1}\sum_{n = 0}^\infty { 1 \over n!} {\rm Tr} \ P \left( i g \oint  t^a A^a_\mu dx_\mu \right)^n,
\eeq   
in terms of connected and disconnected graphs.   Let us set to zero all
connected graphs that have three or more legs, so the Wilson loop is expressed
as a sum of products of temporal and spatial gluon propagators $D_{\mu \nu}(x -
y)$.  We shall show that under this simplifying assumption, (i) the spatial
gluon propagators do not contribute to the Wilson loop, so only the
color-Coulomb potential contributes to the forces on the Wilson loop, and (ii)
the Wilson loop has the value   
\beq
\label{valueofWilson}
W = \exp\left( - g^2 C \int dx_0 \ V [R(x_0)] \right) ,
\eeq
where $R(x_0)$ is the width of the Wilson loop as a function of Euclidean time.
\begin{figure}
        \centering
        \includegraphics[width=6cm]{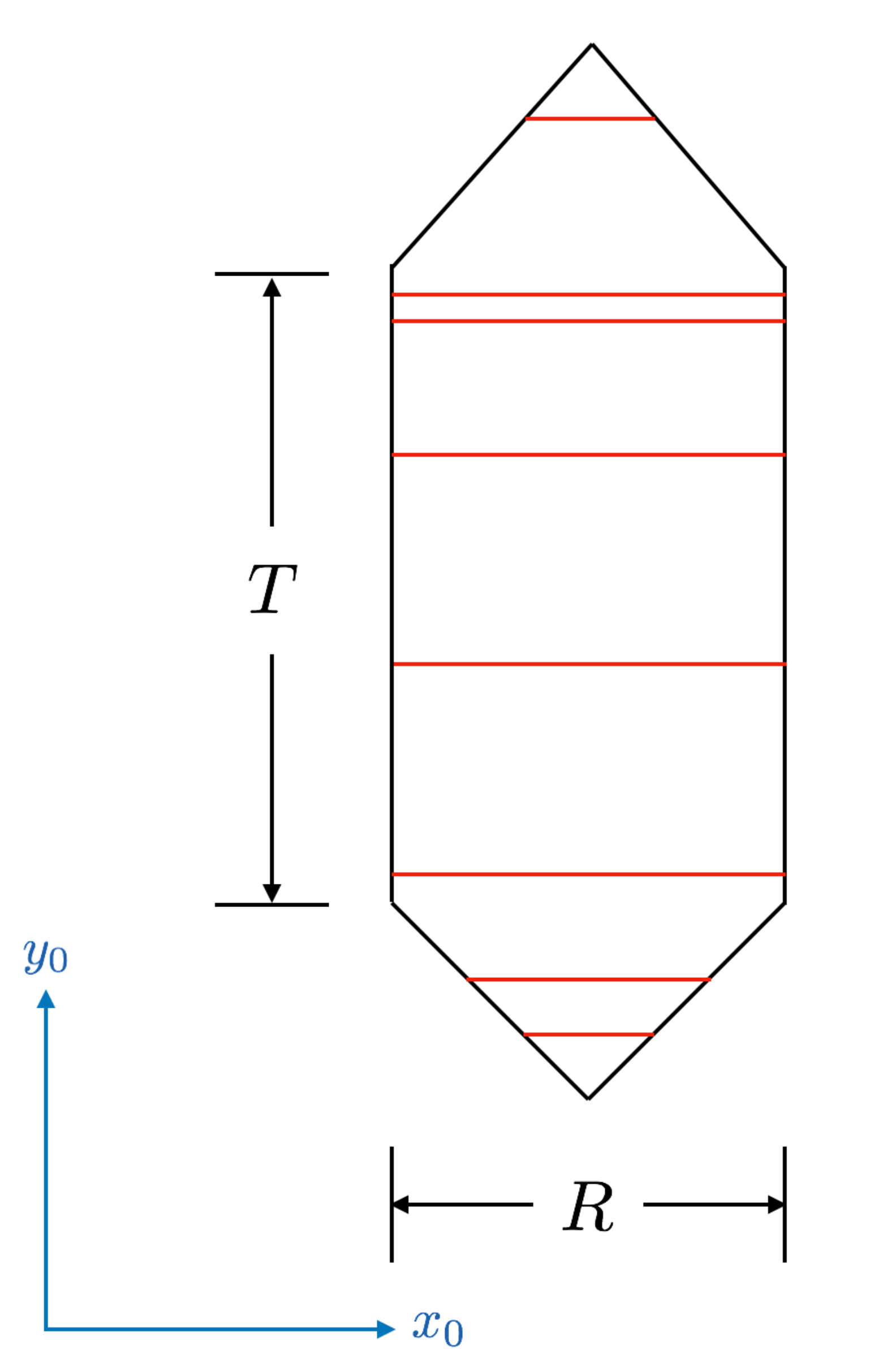}
        \caption{Temporal Gluon propagators form the rungs of Wilson loop, ladder diagrams.}
        \label{wilson_lines}
\end{figure}
This situation is illustrated in Fig.\ \ref{wilson_lines}, where all diagrams are ladder
diagrams, and the (Euclidean) time direction is upward.  This formula is the
same as in an Abelian gauge theory, apart from the Casimir.  Moreover it
implies that the contribution from the parallel parts of the Wilson loop (if
any) is $W_{\rm parallel} = \exp[- g^2 C V(R)T]$, where the parallel parts are
a distance $R$ apart for a time $T$.  

Proof:  Recall that the temporal and spatial propagators vanish away from the
time slice $x_0 - y_0 = 0,$
\beq
D_{\mu \nu}(x - y) = 0 \hspace{1cm} {\rm for} \hspace{1cm} x_0 - y_0 \neq 0, 
\eeq
but the temporal propagator $D_{A_0 A_0}(x-y) = \delta(x_0 - y_0) V({\bf x -
y})$ has the factor $\delta(x_0 - y_0)$, whereas the spatial propagator $D_{\bf
A_i A_j}(x - y)$ is finite at $x_0 - y_0 = 0$, $D_{\bf A_i A_j}(0, \ {\bf x-y)}
= D_{\bf A_i A_j}^{ET}({\bf x - y})$.  Consequently, when the line integral
$\int dx_\mu D_{A_\mu A_\nu}(x_0 - y_0, {\bf x - y} )$ at fixed $y$, crosses
the time-slice $x_0 = y_0$, it receives a finite contribution if the gluon
propagator is temporal ($\mu = 0$), but it receives no contribution when the
gluon propagator is spatial ($\mu = 1, 2, 3$).\footnote{Here we suppose that
the Wilson loop does not have a portion that lies within a time slice. This is
the general case; the Wilson loop can be a circle or an ellipse or an upright
diamond etc.}  Thus the spatial gluon propagators do not contribute to the
Wilson loop, as asserted.  The remaining temporal propagators form the
horizontal rungs of ladder diagrams, as illustrated in Fig.\ref{wilson_lines}.  The
path-ordering makes the two ends of the lowest rung adjacent to each other so,
for the lowest rung, the Lie algebra gives $t^a \delta^{ab} t^b = C_F$, where
$C_F$ is the Casimir in the fundamental representation.  Since it is
proportional to the identity matrix, it may be removed from the path ordering.
The same is then true for the next lowest rung etc., so path ordering gives a
factor of $C_F^{n/2}$.  The combinatorics are then such that the propagators
exponentiate exactly,\footnote{The $n$th term in the power series expansion has
$n![2^{n/2}(n/2)!]^{-1}$ possible pairings when $n$ is even and vanishes when
$n$ is odd.} and we  obtain
\beq
W = \exp\left[ - \half g^2 C_F \oint dx_\mu \delta_{\mu 0} \oint dy_\nu \delta_{\nu 0} \ \delta(x_0 - y_0) V({\bf x - y}) \right]
\eeq
from which \eqref{valueofWilson} follows, as asserted.  The important point is
that the spatial gluon propagators have dropped out, and only the color-Coulomb
potential contributes to the force on the Wilson loop, in this approximation where we have neglected all connected subgraphs with three or more legs.

In Lorentz-covariant gauges, such as the Landau gauge, the gluon propagator is
not instantaneous, and whereas in Coulomb gauge the propagators form the
horizontal rungs of a ladder and the path ordering is easily evaluated, as we
have just seen, in a Lorentz-covariant gauge, the would-be rungs run every which
way, and one does not know how to disentangle the path ordering.  For this
problem, calculation in the Coulomb gauge is simpler than in a
Lorentz-covariant gauge.

\section{Width of the non-instantaneous gluon propagator vanishes in the infrared limit}
\label{app:width_calc}

Suppose the critical exponents satisfy $\gamma = 0$ and $\delta = d + 1$. Then by \eqref{AAc} and \eqref{dspiepiec} we have
\beqa
\Gamma_{\bf AA} = c_1 | {\bf k } |^{d - \delta}  = c_1 | {\bf k } |^{-1}
\nonumber \\
\Gamma_{\tau \tau} = c_2 | {\bf k } |^{d - \delta}  = c_2 | {\bf k } |^{-1},
\eeqa
where $c_1 > 0$ and $c_2 > 0$, which gives, by \eqref{anotherinvert},
\begin{equation}
\label{anotherinverta}
\left(
\begin{array}{rlc}
        D_{{\bf \tau \tau}}(k_0, {\bf k}) & \ D_{{\bf \tau A}}(k_0, {\bf k})   \\
        D_{{\bf A \tau}}(k_0, {\bf k})  & \  D_{{\bf A} {\bf A}}(k_0, {\bf k})
\end{array}
\right)
= { 1 \over k_0^2 + c^2/ {\bf k }^2 }
\left(
\begin{array}{rlc}
 c_1 /  | {\bf k } | \ & \ \ \ \ k_0   \\
        - k_0 \ \ \   & \ \ c_2 / | {\bf k } |
\end{array}
\right),
\end{equation}   
where $c \equiv (c_1 c_2)^{1/2}$.  Upon taking the fourier transform, we obtain
\begin{equation}
\label{anotherinverta}
\left(
\begin{array}{rlc}
        D_{{\bf \tau \tau}}(t, {\bf k}) & \ D_{{\bf \tau A}}(t, {\bf k})   \\
        D_{{\bf A \tau}}(t, {\bf k})  & \  D_{{\bf A} {\bf A}}(t, {\bf k})
\end{array}
\right)
= { 1 \over 2 }
\left(
\begin{array}{rlc}
 (c_1/c_2)^{1/2} \ & \ \ \ \ {\rm sgn}(t)   \\
        - {\rm sgn}(t) \ \ \   & \ \ \ (c_2/c_1)^{1/2}
\end{array}
\right) \exp(- c | t | / | {\bf k}| ).
\end{equation}   
By comparison with $ \langle E| \exp(- H t ) | E \rangle = \langle E| \exp(- E t ) | E \rangle$, we see that the state of a gluon of momentum ${\bf k}$ has energy
\beq
E = c / | {\bf k} |. 
\eeq

Our power-law Ansatz is valid only in the infrared limit ${\bf k} \to 0$.  The
transverse gluon propagator has a peak of width $w \sim | {\bf k} | /c$, which
vanishes in the infrared limit, $ w \to 0$, which corresponds to an
instantaneous propagator at ${\bf k = 0}$. This resolves the apparent paradox
mentioned in section \ref{sec:eqtprops}, that the time dependence of the
propagators violate the remnant symmetries of Coulomb gauge.  While our
approximation scheme seems to break this symmetry, any inferences we make about
the infrared dynamics will be free of this concern since the symmetry is
restored at long distances.



\bibliography{sdelattv3}
\end{document}